\documentclass[preprint,aps,epsfig]{revtex4}

\usepackage{graphicx}
\usepackage{dcolumn}
\usepackage{bm}
\usepackage{epsfig}

\begin{document}

\title{New Method to Calculate Electrical Forces Acting on
a Sphere in an Electrorheological Fluid}
\author{Kwangmoo Kim}
\email{kwangmoo@mps.ohio-state.edu}
\author{David Stroud}
\email{stroud@mps.ohio-state.edu}
\affiliation{Department of
Physics, The Ohio State University, Columbus, Ohio 43210}

\author{Xiangting Li}
\email{xtli@sjtu.edu.cn}
\affiliation{School of Physics and
Astronomy, Raymond and Beverly Sackler Faculty of Exact Sciences,
Tel Aviv University, IL-69978 Tel Aviv, Israel; and Institute of
Theoretical Physics,\nolinebreak Shanghai Jiaotong University,
Shanghai 200240, PR China}

\author{David J. Bergman}
\email{bergman@post.tau.ac.il}
\affiliation{School of Physics and
Astronomy, Raymond and Beverly Sackler Faculty of Exact Sciences,
Tel Aviv University, IL-69978 Tel Aviv, Israel}

\date{\today}


\begin{abstract}

We describe a method to calculate the electrical force acting on a
sphere in a suspension of dielectric spheres in a host with a
different dielectric constant, under the assumption that a spatially
uniform electric field is applied.  The method uses a spectral
representation for the total electrostatic energy of the composite.
The force is expressed as a certain gradient of this energy, which
can be expressed in a closed analytic form rather than evaluated as
a numerical derivative.   The method is applicable even when both
the spheres and the host have frequency-dependent dielectric
functions and nonzero conductivities, provided the system is in the
quasistatic regime. In principle, it includes all multipolar
contributions to the force, and it can be used to calculate
multi-body as well as pairwise forces.  We also present several
numerical examples, including host fluids with finite
conductivities.  The force between spheres approaches the
dipole-dipole limit, as expected, at large separations, but departs
drastically from that limit when the spheres are nearly in contact.
The force may also change sign as a function of frequency when the
host is a slightly conducting fluid.

\end{abstract}



\maketitle

\section{\label{sec:level1}Introduction}

An electrorheological (ER) fluid is a material whose viscosity
changes substantially with the application of an electric
field\cite{hao}. Generally, such fluids are suspensions of
spherical inclusions of dielectric constant
$\epsilon_{\mathrm{i}}$ in a host fluid of a different dielectric
constant $\epsilon_{\mathrm{h}}$. The viscosity is believed to
change because the spheres acquire electric moments (dipole and
higher) when an electric field is
applied, then move under the influence of the electrical forces
between these induced moments. These forces typically cause the
spheres to line up in long chains parallel to the applied field,
thereby increasing the viscosity of the suspension.  The viscosity
relaxes to its usual value when the field is turned off, and the
chain-like structure disappears.

ER fluids have potential applications as variable viscosity fluids
in automobile devices\cite{hartsock}, vibration
control\cite{stanway}, and elsewhere. Furthermore, their operating
principle is also relevant to other materials, such as
magnetorheological (MR) fluids\cite{mr}.  These are suspensions of
magnetically permeable spheres in a fluid of different
permeability, whose viscosity can be controlled by an applied
magnetic field.

To obtain a quantitative theory of ER (and MR) fluids, one needs
to understand the electric-field-induced force among the spheres.
At
low sphere concentrations and large inter-sphere separations, this
force is just that between two interacting electric dipoles whose
magnitude is that of a single sphere in an external electric
field. But at smaller separations, the force deviates from the
dipole-dipole form.  Besides this electrostatic interaction
between the spheres, there are other forces acting on the spheres,
including a viscous frictional force from the host fluid, and a
hard-sphere force when the two dielectric spheres come in contact.
In the present paper, we will be concerned only with the
electrostatic force.

A number of existing theories go beyond the dipole-dipole
approximation in calculating electrostatic forces in ER
fluids\cite{klingenberg,chen,davis1,davis2,khusid,tao,tang1,tang2,clercx},
and several experiments have been carried out which are relevant
to forces in the non-dipole regime (see, e.\ g., Refs.\
\cite{wen,zhiyong}). Klingenberg {\it et al.}\cite{klingenberg}
have incorporated both multipole and multi-body effects into the
sphere-sphere interactions, using a perturbation analysis.  Chen
{\it et al.}\cite{chen} have described a multipole expansion for
the forces acting on one sphere in a chain of spheres in a fluid
of different dielectric constant, and find a strong departure from
the dipolar limit when the particles are closer than about one
diameter. Davis\cite{davis1} has calculated the electrostatic
forces between dielectric spheres in a host fluid directly, using
a finite element approach to solve Laplace's equation for a chain
of particles in a host dielectric.  In a more recent
work\cite{davis2}, he has used an integral equation approach to
calculate the interparticle forces in ER fluids, including effects
due to time-dependent application of an external field, and
nonlinear fluid conductivity. A finite-element approach has also
been used by Tao {\it et al.}\cite{tao} to solve Laplace's
equation and obtain the electrostatic interactions between
particles in a chain of dielectric spheres in a host fluid; they
found, as in Ref.\ \cite{davis1}, that the dipole-dipole
approximation is reasonably accurate for large separations or
moderate dielectric mismatches, but fails in closely spaced
particles and large mismatches. Clercx and Bossis\cite{clercx}
have gone beyond the approximation of dipolar interactions to
include multipolar and many-body interactions, expressed in terms
of the induced multipole moments on each sphere; they also obtain
an expression for the forces in terms of these induced multipole
moments.

As discussed further below, the electrical force acting on a
sphere in an electrorheological fluid is basically the gradient of
the total electrostatic energy of that fluid with respect to the
position of the sphere.  This total electrostatic energy can, in
turn, be expressed in terms of the effective dielectric tensor of
the suspension, a quantity which has been studied since the time
of Maxwell.  Indeed, numerous authors have calculated this tensor
in a wide variety of geometries, going well beyond the regime of
purely dipolar interactions.  For example, Jeffrey\cite{jeffrey}
has calculated the total energy of two spheres in a suspension as
a function of their separation and the dielectric mismatch.  From
this total energy, the force can be obtained numerically as the
derivative of this energy with respect to separation.   Recently,
the pairwise forces between spheres of different sizes have been
calculated using the so-called dipole-induced-dipole
approximation, and even approximately including the effects of
other spheres\cite{yu}.   Once again, the forces were obtained
explicitly by numerically differentiating the total electrostatic
energy with respect to particle coordinates.  McPhedran and
McKenzie\cite{mandm}, and Suen {\it et al.}\cite{suen}, and many
others, have calculated the total energy of spheres arranged in a
periodic structure.  In principle, forces could also be extracted
from this calculation by taking numerical derivatives, provided
that the distortions of the structure leave it periodic. The
energy of a non-periodic suspension of many spheres has been
studied by G\'{e}rardy and Ausloos\cite{ga} and by Fu {\it et
al.}\cite{fu}, in both cases including large numbers of
multipoles. Once again, forces can be extracted, in principle,
from these calculations by taking numerical derivatives of the
computed total energies with respect to sphere coordinates.

Several authors have included the effects of finite conductivity
on forces in electrorheological fluids, and have also considered
how such forces depend on frequency.  Davis\cite{davis3} has
analyzed polarization forces and related effects of conductivity
in ER fluids.  Tang {\it et al.}\cite{tang1,tang2} have calculated
the attractive force between spherical dielectric particles in a
conducting film. Khusid and Acrivos\cite{khusid} have considered
electric-field-induced aggregation in ER fluids, including
interfacial polarization of the particles, the conductivities of
both the particles and the host fluid, and dynamics arising from
dielectric relaxation.  Claro and Rojas\cite{claro} have
calculated the frequency-dependent interaction energy of
polarizable particles in the presence of an applied laser field
within the dipole approximation; they considered primarily optical
frequencies rather than the low frequencies more characteristic of
ER fluids. Ma {\it et al.}\cite{ma} have considered several
frequency-dependent properties of ER systems, starting from a
well-known spectral
representation\cite{berg2,bergman,BergJPC79,bs,BergSiam93,BergDunn92}
for the dielectric function of a two-component composite medium.
Finally, Huang\cite{huang} has carried out a calculation of the
force acting in electrorheological {\em solids} under the
application of a non-uniform electric field, and considering both
finite frequency and finite conductivity effects.

A common feature of most of the above approaches is that they
involve first calculating the {\em total electrostatic energy} of
the suspensions, then obtaining the forces by numerically
differentiating this energy with respect to a particle coordinate.
This numerical differentiation is cumbersome and can be
inaccurate.  Ref.\ \cite{clercx} does give an expression for the
force, but in terms of implicitly defined multipoles. In this
paper, by contrast, we describe a method for calculating these
forces {\em explicitly}, {\em without numerical differentiation}.
This approach is computationally much more accurate than
numerically differentiating the energy.  While our new method may
appear to be merely a computational advance, its additional
accuracy and flexibility should make it widely applicable.

Specifically, our approach allows one to calculate the
electric-field-induced force between two dielectric spheres in a
host of a different dielectric constant, at any separation.   It
is applicable, in principle, to spheres of unequal sizes, to
particles of shape other than spheres, to suspensions in which
either the particle or the host or both have nonzero
conductivities, and to systems whose constituents have
frequency-dependent complex dielectric functions. It can also be
used to calculate the electrostatic force on one particle which is
part of a {\em many-particle} system, and thus is not limited to
two-body interaction. It should thus be useful in quite general
circumstances including, in particular, non-dilute suspensions.

Our approach starts, as do previous calculations, with a method for
calculating the total electrostatic energy of a suspension of (two
or more) spheres in a host material of different dielectric
constant.  We choose to express this total energy in terms of a
certain pole spectrum arising from the quasistatic resonances of the
multi-sphere
system\cite{berg2,bergman,BergJPC79,bs,BergSiam93,BergDunn92}. This
representation has previously been used to calculate the
frequency-dependent shear modulus, static yield stress, and
structures of certain ER systems\cite{ma,thomas,tao2,tao3}. The
force on a given sphere in a multi-sphere system involves a gradient
of this energy with respect to the position of that sphere. But
rather than evaluating this derivative numerically, as in previous
work\cite{ma}, we express this derivative in closed analytical form
in terms of the pole spectrum and certain matrix elements involving
the resonances.  This expression is readily evaluated simply by
diagonalizing a certain matrix, all of whose components are readily
computed.

Our approach has formal similarities to the well-known
Hellmann-Feynman expression for forces in quantum-mechanical
systems\cite{hf}.  In the quantum-mechanical case, the force is
expressed as the negative gradient of system energy with respect
to an ionic position.  This energy is the expectation value of the
Hamiltonian in the ground state.  According to the
Hellmann-Feynman theorem, the gradient operator can be moved
inside the matrix element, thereby eliminating the need to take
numerical derivatives. The Hellmann-Feynman force expression is
the basis for many highly successful molecular dynamics studies in
quantum systems (see, e.\ g., Ref.\ \cite{kresse}).  In the
present classical case, the total energy can also be expressed as
a certain matrix element of an operator, and thus, just as in the
quantum problem, the force is the gradient of that expectation
value.  In this paper, we shall show that, again as in the quantum
case, the gradient operator can
be  moved to within the matrix element, and the need to take a
numerical derivative is eliminated.  This simplification allows,
in principle, the calculation of forces in very complicated
geometries, even though, in the present paper we shall give only
relatively simple numerical illustrations involving forces between
two spherical particles.

The remainder of this paper is organized as follows. In Section
\ref{sec:level2}, we present the formalism necessary to calculate
the forces in a system of two or more dielectric spheres in a host
medium, without taking a numerical derivative.  In Section
\ref{sec:level3}, we give several numerical examples of these
forces, at both zero and finite frequencies, for a two-sphere
system.  Section \ref{sec:level4} presents a concluding discussion
and suggestions for future work.

\section{\label{sec:level2}Formalism}

Let us assume that we have a composite consisting of spherical
inclusions of isotropic dielectric constant
$\epsilon_{\mathrm{i}}$ in a host of
isotropic dielectric constant $\epsilon_{\mathrm{h}}$, both of
which may be complex and frequency-dependent. We will assume that
a spatially uniform electric field $\mathrm{Re}\!\left[\,{\bf
E}_0\,e^{-i\omega t}\right]$ is applied in an arbitrary direction
(we take ${\bf E}_0$ real).  We also assume that the system is in
the ''quasistatic regime.''  In this regime, the product $k\,\xi
\ll 1$, where $k$ is the wave vector and $\xi$ is a characteristic
length scale describing the spatial variation of $\epsilon({\bf
x},\omega)$. Under these conditions, the local electric field
${\bf E}({\bf x},\omega)=-{\bf\nabla}\Phi$, where $\Phi$ is the
electrostatic potential.  Finally, we assume that the ${\bf
R}^{th}$ spherical inclusion is centered at ${\bf R}$, and has
radius $a_{\bf R}$.  The approach which we use automatically
includes all local field effects.

Since our force expressions differ slightly at zero and finite
frequencies, we will first present the formalism at $\omega = 0$,
and then generalize the results to finite $\omega$.

\vspace{0.1in}

\noindent {\bf A. Zero Frequency}

\vspace{0.1in}

If the position of the spheres is fixed, the total electrostatic
energy may be written in the form
\begin{equation}
W=\frac{V}{8\,\pi}\sum_{i=1}^3\sum_{j=1}^3\epsilon_{\mathrm{e};ij}\,
E_{0,i}\,E_{0,j}, \label{eq:w}
\end{equation}
where $V$ is the system volume, $\epsilon_{\mathrm{e};ij}$ is a
component of the
macroscopic effective dielectric tensor, and $E_{0,i}$ is a
component of the applied electric field.  Eq.\ (\ref{eq:w}) is, in
fact, a possible definition of
$\epsilon_{\mathrm{e};ij}(\omega)$\cite{bs}. To produce this
applied field, we require that $\Phi({\bf x}) = -{\bf
E}_0\cdot{\bf x}$ at the boundary $S$ of the system, which is
assumed to be a closed surface enclosing $V$.  In writing eq.\
(\ref{eq:w}), we allow for the possibility that the spheres in the
composite are arranged in such a way that the composite is
anisotropic even though its components are not.

For an {\em isotropic} composite, $\epsilon_{\mathrm{e}}$ may be
written in terms of a certain pole spectrum of the composite
as\cite{berg2,bergman,BergJPC79,bs,BergSiam93,BergDunn92}
\begin{equation}
1 - \frac{\epsilon_{\mathrm{e}}}{\epsilon_{\mathrm{h}}} =
\sum_\alpha\frac{B_\alpha}{s - s_\alpha}, \label{eq:epse}
\end{equation}
where
\begin{equation}
s = \frac{1}{1 - \epsilon_{\mathrm{i}}/\epsilon_{\mathrm{h}}},
\label{eq:s}
\end{equation}
$s_\alpha$ is a pole, and $B_\alpha$ is the corresponding residue.
The poles $s_\alpha$ are confined to the interval $0\leq s_\alpha <
1$.  For an {\em anisotropic} composite, this form may be
generalized to
\begin{equation}
\delta_{ij}
-\frac{\epsilon_{\mathrm{e};ij}}{\epsilon_{\mathrm{h}}} =
\sum_\alpha\frac{B_{\alpha;ij}}{s - s_\alpha},
\label{eq:epseij}
\end{equation}
where $\delta_{ij}$ is a Kronecker delta function and
$B_{\alpha;ij}$ is a matrix of residues. This form is general,
applicable to any two-component composite material which is made
up of isotropic constituents, but is not necessarily isotropic
macroscopically.  As in the isotropic case, the poles are confined
to the interval $0\leq s_\alpha < 1$.

The poles $s_\alpha$ are the eigenvalues of a certain Hermitian
operator $\Gamma$, and the residues $B_\alpha$ are determined by the
eigenvectors of that operator. $\Gamma$ is defined in terms of its
operation on an arbitrary function $\phi({\bf r})$ by the relation
\begin{equation}
\Gamma\phi({\bf r}) \equiv \int_{v_{\mathrm{tot}}} \mathrm{d}^3r^\prime
\,{\bf
\nabla}^\prime\!\! \left( \frac{1}{|{\bf r}-{\bf
r}^\prime|}\right) \cdot{\bf \nabla}^\prime\!\phi({\bf r}^\prime),
\label{eq:integro}
\end{equation}
where the integration runs over the total volume $v_{\mathrm{tot}}$
of all the inclusions of $\epsilon_{\mathrm{i}}$.   As in Ref.\
\cite{KantorBerg82}, we introduce a ``bra-ket'' notation for two
potential functions to denote their inner product,
\begin{equation}
\langle \phi |\psi\rangle \equiv \int_{v_{\mathrm{tot}}}
\mathrm{d}^3r \,{\bf \nabla}\phi^*\!({\bf r})\cdot{\bf
\nabla}\psi({\bf r}). \label{eq:inner}
\end{equation}
Physically, the eigenvalues correspond to the frequencies of the
natural electrostatic modes of the composites, at which charge can
oscillate without any applied field, and the corresponding
eigenvectors describe the electric fields of those modes.

It is convenient to express $\Gamma$ in terms of its matrix elements
between the normalized eigenstates of isolated spheres.  In this
basis, and using eqs.\ (\ref{eq:integro}) and (\ref{eq:inner}), it
is found that $\Gamma$ has the following matrix elements:
\begin{equation}
\Gamma_{{\bf R} \ell m; {\bf R}^\prime \ell^\prime m^\prime} =
\langle \psi_{{\bf R} \ell m}|\Gamma \,
 \psi_{{\bf R}^\prime \ell^\prime m^\prime}\rangle =
s_\ell \,\delta_{\ell,\ell^\prime}\delta_{m,m^\prime}\delta_{{\bf
R},{\bf R}^\prime} +Q_{{\bf R}\ell m; {\bf R}^\prime \ell^\prime
m^\prime}(1 - \delta_{{\bf R},{\bf R}^\prime}), \label{eq:gamma}
\end{equation}
where $s_\ell$ and $Q_{{\bf R}\ell m; {\bf R}^\prime\ell^\prime
m^\prime}$ will be given further below.  Inside the sphere centered
at {\bf R}, $\psi_{{\bf R} \ell m}({\bf r})$ is equal to an
eigenfunction or resonance state of that isolated sphere, while
outside that sphere $\psi_{{\bf R} \ell m}({\bf r})=0$. The angular
dependence of $\psi_{{\bf R} \ell m}({\bf r})$ is given by the
spherical harmonic $Y_{\ell m}(\theta,\phi)$, which has an order
$\ell m$ multipole moment of electric polarization. However, the
eigenvalue of this state depends only on $\ell$:
\begin{equation}
s_{{\bf R} \ell m}=s_\ell=\frac{\ell}{2\,\ell+1}.
\end{equation}

The quantity $Q_{{\bf R}\ell m;{\bf R}^\prime\ell^\prime
m^\prime}$ represents the matrix element of $\Gamma$ between two
states of two different, non-overlapping spheres (i.\ e., $|{\bf
R}^\prime-{\bf R}|
> a_{\bf R}+a_{{\bf R}^\prime}$) and is given by
\begin{eqnarray}
Q_{{\bf R}\ell m;{\bf R}^\prime \ell^\prime m^\prime}& = &
(-1)^{\ell^\prime + m^\prime} \frac{a_{\bf R}^{\ell+1/2}a_{\bf
R^{\prime}}^{\ell^{\prime}+1/2}}{|{\bf R}^{\prime}-{\bf
R}|^{\ell+\ell^\prime + 1}}\left(\frac{\ell\,\ell^\prime}
{(2\,\ell+1)(2\,\ell^\prime + 1)}\right)^{\!\!1/2} \nonumber \\
& & \times \frac{(\ell+\ell^\prime + m -
m^\prime)!}{\left[(\ell+m)!(\ell-m)!
(\ell^\prime+m^\prime)!(\ell^\prime-m^\prime)!\right]^{1/2}}
\nonumber
\\ & & \times \, e^{i\,\phi_{{\bf R}^\prime-{\bf R}}(m^\prime - m)}
P_{\ell^\prime +\ell}^{\,m^\prime - m} (\cos\theta_{{\bf
R}^\prime-{\bf R}}), \label{eq:q}
\end{eqnarray}
where $\theta_{{\bf R}^\prime - {\bf R}}$ and $\phi_{{\bf
R}^\prime-{\bf R}}$ are polar and azimuthal angles of the vector
${\bf R}^\prime - {\bf R}$, and the functions
$P_{\ell^\prime+\ell}^{\,m^\prime-m}$ are the associated Legendre
polynomials.

If we denote the eigenfunctions of $\Gamma$ by $\psi_\alpha({\bf
r})$, then $s_\alpha$ and $\psi_\alpha({\bf r})$ satisfy the
eigenvalue equation
\begin{equation}
\Gamma\,\psi_\alpha({\bf r}) = s_\alpha\psi_\alpha({\bf r}).
\label{eq:eigen}
\end{equation}
Since $\Gamma$ is a Hermitian operator, the eigenvalues $s_\alpha$
are real, and the corresponding eigenfunctions are orthogonal and
can be chosen to be orthonormal.  Again, it is convenient to
represent them using a bra-ket notation.  In this notation, the
eigenfunctions are denoted $|\alpha\rangle$ and the orthonormality
condition is
\begin{equation}
\langle \alpha|\alpha^\prime\rangle = \delta_{\alpha,\alpha^\prime}.
\end{equation}
The eigenvalues $s_\alpha$ are the poles of eq.\ (\ref{eq:epse})
or eq.\ (\ref{eq:epseij}).

The corresponding residues $B_{\alpha,ij}$ may be expressed in the
same bra-ket notation as
\begin{equation}
B_{\alpha;ij} = \frac{v_{\mathrm{tot}}}{V}\langle
i|\alpha\rangle\langle\alpha|j\rangle \equiv
M_{\alpha}^{\,i}\,M_{\alpha}^{\,j\,\ast}, \label{eq:balpha}
\end{equation}
The matrix element $M_{\alpha}^{\,i} = \langle i|\alpha\rangle$ is
basically the component of the electric dipole moment of the
eigenfunction $|\alpha\rangle$ in the $i^{th}$ Cartesian
direction.

It is convenient to expand both the eigenfunctions
$|\alpha\rangle$ and the states $|i\rangle$ ($i = x$, $y$, $z$),
in terms of the single-sphere eigenfunctions $\psi_{{\bf R} \ell
m}({\bf r})$ mentioned above. In bra-ket notation,
\begin{equation}
|\alpha\rangle = \sum_{{\bf R}\ell m}A_{{\bf R}\ell m}^\alpha|{\bf
R}\ell m\rangle.
\end{equation}
The expansion coefficients satisfy the normalization condition
$\sum_{{\bf R}\ell m}|A_{{\bf R}\ell m}^\alpha|^2 = 1$, where the
indices $\ell = 1, 2, \ldots$ and $m = -\ell, -\ell+1, \ldots,
+\ell$ respectively.  Similarly, the states $|i\rangle$ may be
expanded as
\begin{equation}
|i\rangle = \sum_{{\bf R}\ell m}M_{{\bf R}\ell m}^{\,i}|{\bf
R}\ell m\rangle,
\end{equation}
where $i = x$, $y$, $z$.   If the $z$ axis is chosen as the polar
axis for the spherical harmonics, then the $M_{{\bf R}\ell
m}^{\,i}$ take the form\cite{bergman}
\begin{eqnarray}
M_{{\bf R} \ell m}^{\,x} & = & \left(\frac{v_{\bf
R}}{2\,v_{\mathrm{tot}}}\right)^{\!\!1/2}
(\delta_{m,1}+\delta_{m,-1})\,\delta_{\ell,1}, \nonumber \\
M_{{\bf R}\ell m}^{\,y} & = & -i\left(\frac{v_{\bf
R}}{2\,v_{\mathrm{tot}}}\right)^{\!\!1/2}
(\delta_{m,1}-\delta_{m,-1})\,\delta_{\ell,1}, \nonumber \\
M_{{\bf R}\ell m}^{\,z} & = & \left(\frac{ v_{\bf
R}}{v_{\mathrm{tot}}}\right)^{\!\!1/2}\delta_{m,0}\,\delta_{\ell,1}.
\label{eq:mxyz}
\end{eqnarray}
Thus, the matrix elements $M_{\alpha}^{\,i}$ are given explicitly
by
\begin{eqnarray}
M_{\alpha}^{\,x} & = & \sum_{\bf R}\left(\frac{v_{\bf
R}}{2V}\right)^{\!\!1/2}
(A_{{\bf R}11}^\alpha+A_{{\bf R}1-1}^\alpha),  \nonumber \\
M_{\alpha}^{\,y} & = &  -i\sum_{\bf R}\left(\frac{v_{\bf
R}}{2V}\right)^{\!\!1/2}(A_{{\bf R}1 1}^\alpha -
A_{{\bf R}1-1}^\alpha),\nonumber \\
M_{\alpha}^{\,z} & = & \sum_{\bf R}\left(\frac{v_{\bf
R}}{V}\right)^{\!\!1/2}A_{{\bf R} 1 0}^\alpha . \label{eq:malphi}
\end{eqnarray}
In other words, the residues of the $\alpha^{th}$ eigenfunction are
basically the square of the electric dipole moment of that mode in
the $x$, $y$, or $z$ direction.

Combining these results, we can re-express the matrix elements
(\ref{eq:epseij}) of the dielectric tensor in bra-ket notation first
as
\begin{equation}
\delta_{ij} - \frac{\epsilon_{\mathrm{e};ij}}{\epsilon_{\mathrm{h}}}
= \frac{v_{\mathrm{tot}}}{V}\sum_{\alpha}\frac{\langle
i|\alpha\rangle \langle\alpha|j\rangle}{s - s_\alpha},
\label{eq:epse1}
\end{equation}
where the explicit forms of $\langle i|\alpha\rangle$ and $\langle
\alpha|j\rangle$ are given by eqs.\ (\ref{eq:malphi}).

We now use the above formalism to obtain an expression for the force
on a dielectric sphere centered at {\bf R} in a suspension
consisting of an arbitrary assembly of spheres.  First, we rewrite
eq.\ (\ref{eq:epse1}) as
\begin{equation}
\delta_{ij} -
\frac{\epsilon_{\mathrm{e};ij}}{\epsilon_{\mathrm{h}}} =
\frac{v_{\mathrm{tot}}}{V}\langle i|G(s)|j\rangle,
\label{eq:epse2}
\end{equation}
where
\begin{equation}
G(s) \equiv \sum_\alpha\frac{|\alpha\rangle\langle\alpha|} {s -
s_\alpha} = (sI - \Gamma)^{-1}, \label{eq:g}
\end{equation}
is a Green's function for this problem, $I$ is the identity
matrix, and the matrix elements of $\Gamma$ are given by eq.\
(\ref{eq:gamma}).   If the applied electric field is ${\bf E}_0$,
the total energy takes the form
\begin{equation}
W = \frac{V}{8\,\pi}\,\epsilon_{\mathrm{h}}\,{\bf E}_0\cdot{\bf
E}_0
-\frac{\,v_{\mathrm{tot}}\epsilon_{\mathrm{h}}}{8\,\pi}\sum_{ij}E_{0,i}\,\langle
i| G(s)|j\rangle\, E_{0,j}. \label{eq:w2}
\end{equation}

We now write the $k^{th}$ component of the force on the sphere at
${\bf R}$ as
\begin{equation}
F_{R_{k}} = +\!\left(\frac{\partial W}{\partial
R_k}\right)_{\!\!\Phi}. \label{eq:force}
\end{equation}
Here $R_k$ denotes the $k^{th}$ component of ${\bf R}$, and the
subscript $\Phi$ denotes that the derivative is taken with the
potential fixed on the boundaries. The positive sign, though
seemingly counterintuitive, is actually correct here because the
system is held at fixed {\em potential} on the
boundaries\cite{jackson}. Using eq.\ (\ref{eq:w2}), the derivative
in eq.\ (\ref{eq:force}) can be expressed as
\begin{equation}
\left(\frac{\partial W}{\partial R_k}\right)_{\!\!\Phi} =
-\frac{\,v_{\mathrm{tot}}\epsilon_{\mathrm{h}}}{8\,\pi}
\sum_{ij}E_{0,i}\,E_{0,j}\langle i|\left(\frac{\partial}{\partial
R_k}G(s)\right)_{\!\!\Phi}|j\rangle. \label{eq:dwdrk}
\end{equation}
The derivative can be brought inside the bras and kets because
these bras and kets do not depend on $R_k$.

The derivative of $G(s)$ appearing in eq.\ (\ref{eq:dwdrk}) can be
evaluated straightforwardly.  Let us assume that the operator
$\Gamma$ depends on some scalar parameter $\lambda$ (e.\ g.,
$R_k$). Then, if we introduce the operator $U_\lambda =
\partial \Gamma/\partial \lambda$, we can calculate the partial
derivative $\partial G(s,\lambda)/\partial \lambda$ as follows:
\begin{eqnarray}
\frac{\partial G(s, \lambda)}{\partial \lambda} & = &
\lim\limits_{\partial \lambda \to 0} \left[\{sI - \Gamma(\lambda)
- U_{\lambda}\,\partial \lambda\}^{-1} - \{sI -
\Gamma(\lambda)\}^{-1}\right]/\,\partial \lambda
\nonumber \\
& = & \lim\limits_{\partial \lambda \to 0} \left[\{I - (sI -
\Gamma(\lambda))^{-1}\,U_{\lambda}\,\partial \lambda\}^{-1}(sI -
\Gamma(\lambda))^{-1} - \{sI -
\Gamma(\lambda)\}^{-1}\right]/\,\partial \lambda
\nonumber \\
& = & [\,sI -\Gamma(\lambda)\,]^{-1}\,U_{\lambda}\,
[\,sI-\Gamma(\lambda)\,]^{-1} \nonumber \\
& = &
G(s,\lambda)\,U_{\lambda}G(s, \lambda).
\label{eq:gug}
\end{eqnarray}

We can now use the above identity to calculate the force as given in
eqs.\ (\ref{eq:force}) and (\ref{eq:dwdrk}). The result is
\begin{equation}
F_{R_k}
=-\frac{\,v_{\mathrm{tot}}\epsilon_{\mathrm{h}}}{8\,\pi}\sum_i\sum_j
E_{0,i}\,E_{0,j}\langle i|G(s, \lambda)\,U_{R_k}G(s,
\lambda)|j\rangle, \label{eq:frk}
\end{equation}
where we have introduced
\begin{equation}
U_{R_k} =\frac{\partial\Gamma}{\partial R_k}.
\end{equation}
Using the representation (\ref{eq:g}) for $G(s,\lambda)$ [and
taking the eigenvalue $s_\alpha$ and the eigenstate
$|\alpha\rangle$ to refer to the operator $\Gamma(R_k)$], we can
rewrite eq.\ (\ref{eq:frk}) as
\begin{equation}
F_{R_k} =-\frac{\,v_{\mathrm{tot}}\epsilon_{\mathrm{h}}}{8\,\pi}
\sum_i\sum_jE_{0,i}\,E_{0,j}\sum_\alpha\sum_\beta \frac{\langle
i|\alpha\rangle\langle\alpha|U_{R_k}|\beta\rangle\langle\beta|j\rangle}
{(s-s_\alpha)(s-s_\beta)}.\label{eq:force2}
\end{equation}
Eq.\ (\ref{eq:force2}) is our central formal result.

As noted earlier, eq.\ (\ref{eq:dwdrk}) bears a resemblance to the
Hellmann-Feynman theorem in quantum mechanics\cite{hf}: in both
cases, the derivative of an operator with respect to a parameter
appears inside a matrix element. But there is a significant
difference between the two. In the Hellmann-Feynman case, the ket
which plays the role of $|i\rangle$ is an eigenstate of an
operator, which is the actual Hamiltonian of the system. Although
the ket in that case depends on $\lambda$, the derivative can
still be moved inside the bra and ket because the eigenstates are
orthonormalized. Here, by contrast, the states $|i\rangle$ and
$|j\rangle$ are not eigenstates of the operator $\Gamma$, but they
do not depend on $\lambda$; so the derivative can still be moved
inside the matrix element.  This simplification allows forces to
be computed without carrying out numerical derivatives.

Eq.\ (\ref{eq:force2}) may appear to be rather difficult to apply
in practice.  But in fact it is computationally quite tractable.
Basically, there are two matrices which are needed as inputs: $G$
and $U_{R_k} = \partial \Gamma/\partial R_k$.  $G$ is diagonal in
the same basis as $\Gamma$.  All the matrix elements of $\Gamma$
are explicitly known in the ${\bf R}\ell m$ basis [cf.\ eq.\
(\ref{eq:gamma})].  Likewise, the matrix elements of
$\partial\Gamma/\partial R_k$ can be obtained from $\Gamma$ purely
by elementary calculus.  Thus, in order to compute the force
component $F_{R_k}$, one first finds the eigenvalues and
eigenfunctions of $\Gamma$ (and hence of $G$), then computes the
matrix $G\,U_{R_{k}}G$ in the basis in which $\Gamma$ is diagonal,
and finally the matrix elements $\langle
i|G\,U_{R_{k}}G|j\rangle$, from which the force can be computed
for any direction of the applied field ${\bf E}_0$. Since the
diagonalization can be done with standard computer packages, the
whole procedure is well defined and straightforward. Furthermore,
once $\Gamma$ has been diagonalized, the same basis can be used to
compute the forces for any value of the variable $s = (1
-\epsilon_{\mathrm{i}}/\epsilon_{\mathrm{h}})^{-1}$.


To illustrate how this formalism can actually be used to compute the
force explicitly, we will consider just a suspension of two spheres,
the two spheres being
located at $(0, 0, 0)$ and $(0, 0, R_0)$.   The total energy is
given by eq.\ (\ref{eq:w}).  We consider two configurations for the
electric field: ${\bf E}_0 = (0, 0, E_0)$ (we call this the
``parallel configuration'') and ${\bf E}_0 = (E_0, 0, 0)$
(``perpendicular configuration''). In both cases, the component of
the force on the sphere at ${\bf R}_0$ along the axis joining the
two spheres can be calculated using eq.\ (\ref{eq:force2}).

To compute the force explicitly in this example, we have to
consider how the operator $\Gamma$ changes with the separation
$R_0$ of the spheres, so that we can compute the matrix elements
of $U_{R_{0}}$. According to eqs.\ (\ref{eq:gamma}) and
(\ref{eq:q}), the diagonal matrix elements of $\Gamma$ are
independent of $R_0$, while each of the off-diagonal matrix
elements, according to eq.\ (\ref{eq:q}), is proportional to an
integer power of $1/|{\bf R}^{\prime} - {\bf R}| \equiv 1/R_0$.
Hence, $U_{R_{0}} \equiv \partial \Gamma/\,\partial R_0$ is easily
calculated in a closed form. For the case of two spheres, it is
straightforward to calculate this derivative.  The eigenstates
$|\alpha\rangle$, as well as $s_\alpha$, are already known if the
original eigenvalue problem involving $\Gamma(R_0)$ has been
solved. The kets $|i\rangle$ are given by eq.\ (\ref{eq:mxyz}).
Therefore, it is straightforward to calculate the quantity
$\langle \alpha|U_{R_0}|\beta \rangle$ and hence the force, using
eq.\ (\ref{eq:force2}).

\vspace{0.1in}

\noindent {\bf B. Finite Frequencies}

\vspace{0.1in}

The results of the previous subsection are readily generalized to
finite frequencies.  In this case, the total electrostatic energy
will be a sinusoidally varying function of time.  The quantities of
experimental interest will be the time-averaged electrostatic energy
$W_{\mathrm{av}}$ and time-averaged forces. $W_{\mathrm{av}}$ is
given by the generalization of eq.\ (\ref{eq:w}), with an extra
factor of $1/2$ to take into account time-averaging, namely
\begin{equation}
W_{\mathrm{av}} =
\frac{V}{16\,\pi}\,\mathrm{Re}\!\left[\,\sum_{i=1}^3\sum_{j=1}^3
\,\epsilon_{\mathrm{e};ij}(\omega)\,E_{0,i}\,E_{0,j}\,\right].
\end{equation}
Here the applied field is assumed to be ${\bf E}_0\cos(\omega t) =
\mathrm{Re}\!\left[\,{\bf E}_0\,e^{-i\omega t}\right]$,
$\epsilon_{\mathrm{e};ij}(\omega)$ is a component of the complex
frequency-dependent
macroscopic effective dielectric tensor, and ${\bf E}_0$ is a real
vector. All the remaining equations in Sec. \ref{sec:level2}A
continue to be valid up to eq.\ (\ref{eq:force}), which is
replaced by
\begin{equation}
F_{\mathrm{av},R_{k}} = +\!\left (\frac{\partial
W_{\mathrm{av}}}{\partial R_{k}}\right )_{\!\!\Phi}, \label{eq:f1}
\end{equation}
where $\Phi$ is held fixed. The generalization of eq.
(\ref{eq:force2}) is
\begin{equation}
F_{\mathrm{av},R_k} =
-\mathrm{Re}\!\left[\frac{\,v_{\mathrm{tot}}\epsilon_{\mathrm{h}}}
{16\,\pi}\sum_{i=1}^3\sum_{j=1}^3
E_{0,i}\,E_{0,j}\sum_\alpha\sum_\beta\frac{\langle i|\alpha
\rangle \langle \alpha|U_{R_k}|\beta \rangle \langle \beta|j
\rangle}{(s-s_\alpha)(s-s_\beta)}\right]. \label{eq:fcomplex}
\end{equation}
Expression (\ref{eq:fcomplex}) can be evaluated just as at $\omega
= 0$, and thus the time-averaged force at finite frequency can
also be computed explicitly.

\section{\label{sec:level3}Numerical Results}

\begin{figure}
  \hfill
  \begin{minipage}[t]{.47\textwidth}
    \begin{center}
      \epsfig{file=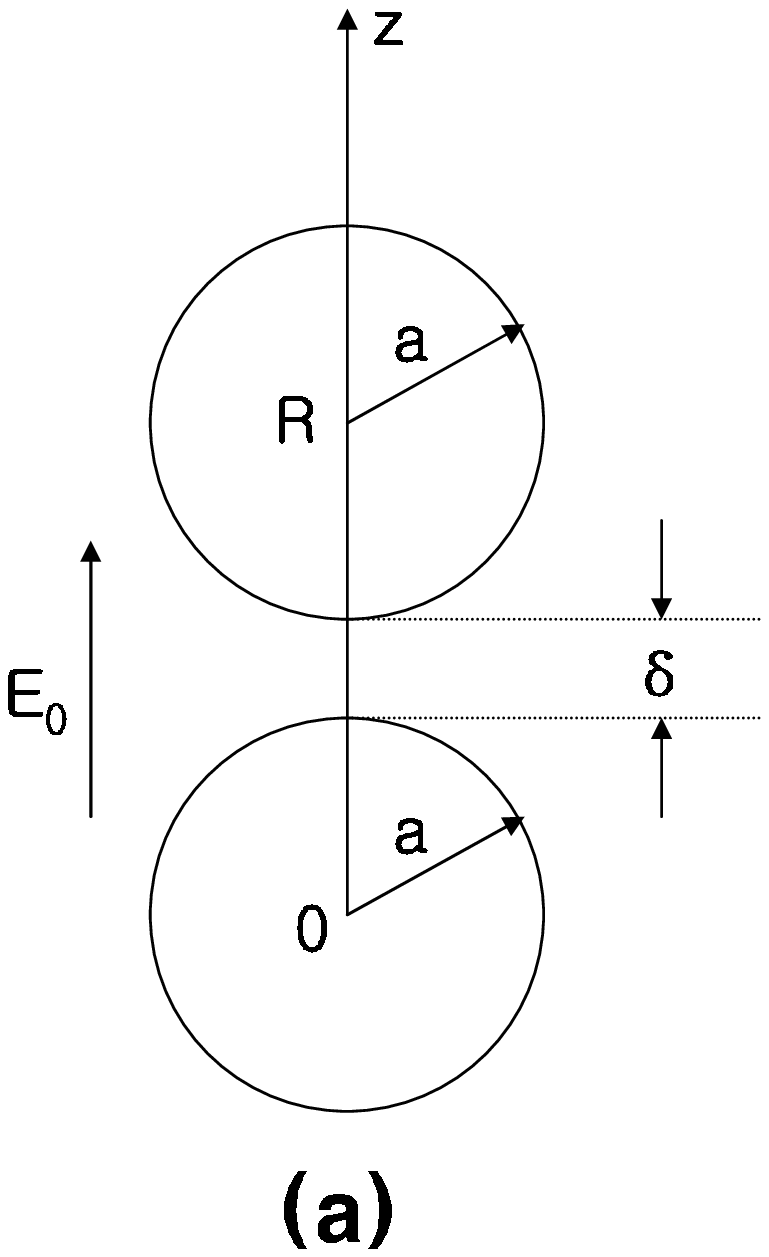, scale=0.6}
    \end{center}
  \end{minipage}
  \hfill
  \begin{minipage}[t]{.47\textwidth}
    \begin{center}
      \epsfig{file=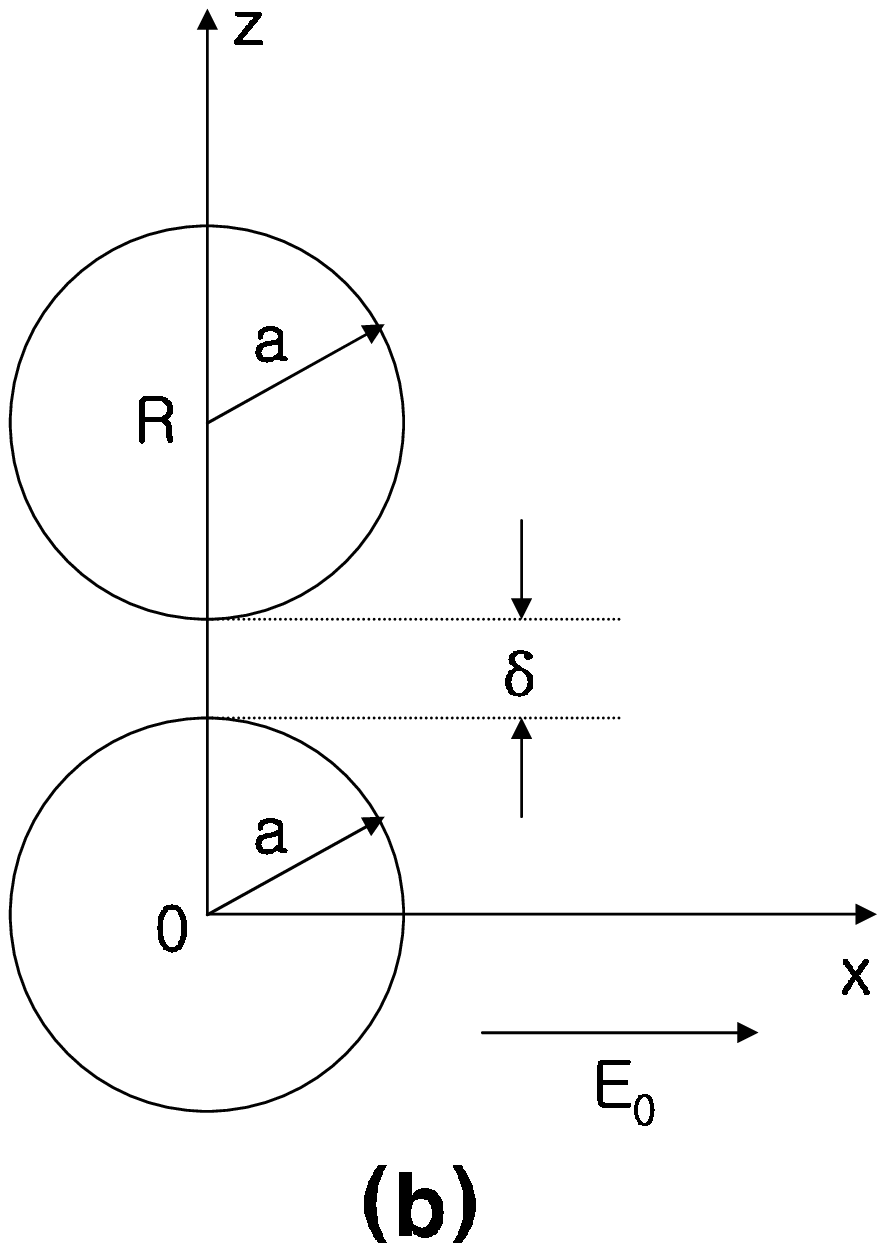, scale=0.6}
    \end{center}
  \end{minipage}
  \hfill
\caption{\label{fig:config} \footnotesize{Geometry considered in
most of our calculations: Two identical spheres of radius $a$ are
located at the origin and at $z=R$, and are contained in a host
material. $\delta$ is the surface-to-surface distance between the
two spheres. The complex dielectric function of the spheres is
$\epsilon_{\mathrm{i}}(\omega)$ and that of the host material is
$\epsilon_{\mathrm{h}}(\omega)$.  A spatially uniform electric
field is applied in the $z$ direction in (a) and in the $x$
direction in (b).}}
\end{figure}

We have applied the above formalism to two spheres of dielectric
constant $\epsilon_{\mathrm{i}}$ in a host of dielectric constant
$\epsilon_{\mathrm{h}}$.  In most cases, we assume that the
spheres have the same radius.  We choose a coordinate system such
that the two spheres are located at the origin and at ${\bf R} =
R\,{\bf \hat{z}}$, and we consider two configurations for the
applied electric field, ${\bf E}_0 = E_0{\bf \hat{z}}$ and ${\bf
E}_0 = E_0{\bf \hat{x}}$, as shown in Fig.\ \ref{fig:config}.

Once the elements of the $\Gamma$ and $U_{R}$ matrices are known,
the calculation of the interparticle force reduces to an
eigenvalue problem.   To carry out the various required matrix and
vector operations, we used GNU Scientific Library (GSL)
routines\cite{note1} and C++ complex class library. In the
parallel configuration, we calculated all the elements in the
$\Gamma$ and $U_{R}$ matrices up to $\ell_{\mathrm{max}}=80$; it
is easy to include such a large cutoff because only $m = 0$ needs
to be considered for this geometry, the polar and azimuthal angles
of ${\bf R}$ equaling zero.  Despite the large cutoff, most of the
contributions to these matrices came from $\ell < 10$.   Based on
this information, we set $\ell_{\mathrm{max}}= 10$ for the
$\Gamma$ and $U_{R}$ matrices in the perpendicular geometry. Even
with this cutoff, the matrices involved in this calculation are
large since $m$ can be nonzero in the perpendicular case: the
dimension of the matrix for $\ell \leq 10$ is
$2\sum_{\ell=1}^{10}(2\,\ell+1)=240$.

The $\Gamma$ matrix for both cases consists of four square blocks.
The two diagonal square blocks have diagonal elements
$s_{\ell}=\ell/(2\,\ell+1)$ with all off-diagonal elements
vanishing. The other two (off-diagonal) square blocks have
elements $Q_{0\ell m;{\bf R}\ell^{\prime}m^{\prime}}$.   For the
$U_{R}$ matrix, the diagonal square blocks have all zero elements,
and the elements of the two off-diagonal blocks are equal to
$\partial Q_{0\ell m;\mathbf{R}\ell^{\prime}m^{\prime}}/
\,\partial R$. Once we have calculated all the eigenvalues and
eigenvectors of the $\Gamma$ matrix, we can compute the
$M_{\alpha}$, $B_{\alpha}$, and hence the force on the sphere from
$\langle \alpha|U_{R}|\beta \rangle$, using eq.\ (\ref{eq:force2})
or (\ref{eq:fcomplex}).

\begin{figure}
\begin{center}
\epsfxsize=0.5\textwidth \epsfbox{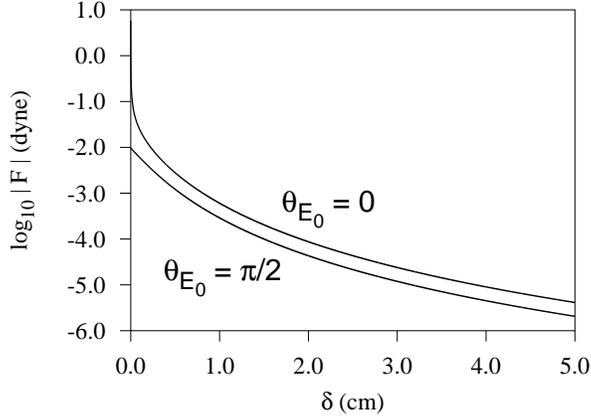}
\end{center}
\caption{\label{fig:eqsize} \footnotesize{Magnitude of the radial
component of the force at zero frequency between two identical
spheres of radius $a$, with $\epsilon_{\mathrm{i}} = 10^5$,
$\epsilon_{\mathrm{h}} = 1$, plotted as a function of sphere
separation, for electric field parallel to axis between spheres
$(\theta_{{\bf E}_{0}}=0)$, and field perpendicular to that axis
$(\theta_{{\bf E}_{0}}=\pi/\,2)$.  Note the logarithmic scale on the
vertical axis.  In both cases, we assume sphere radii of
$3.15\,\mathrm{mm}$, and an electric field of strength
$25.2\,\mathrm{V\!/mm}$, as in Ref.\ \cite{zhiyong}. The force in
the parallel field case is negative (attractive) while that in the
perpendicular field case is positive (repulsive).  In this and the
following two plots, the force is calculated at zero frequency.}}
\end{figure}

As a first example, we have considered $\epsilon_{\mathrm{i}} =
10^5$, $\epsilon_{\mathrm{h}} = 1$. The choice for
$\epsilon_{\mathrm{i}}$ approximates the value
$\epsilon_{\mathrm{i}}=\infty$ corresponding to two metallic
spheres at zero frequency in an insulating host with unit
dielectric constant. In Fig.\ \ref{fig:eqsize}, we show the {\em
magnitude} of the calculated radial component of the force acting
on the sphere at ${\bf R}$, as a function of the sphere
separation, for both parallel and perpendicular configurations.
Although not apparent from the plot, this component of the force
is attractive (i.\ e., negative) in the parallel configuration,
repulsive in the perpendicular configuration.
We have arbitrarily chosen sphere radii of $a = 3.15\,\mathrm{mm}$
and a field strength of $E_0 = 25.2\,\mathrm{V\!/mm}$ as in recent
experiments carried out in Ref.\ \cite{zhiyong} (for different
materials).  However, the forces are easily scaled with both field
strength and sphere radii: for fixed $\epsilon_{\mathrm{i}}$ and
$\epsilon_{\mathrm{h}}$ the appropriate scaling relation is
\begin{equation}
F_{12}^{\perp,\,\|} = a^6E_0^{\,2}f_{\perp,\,\|}
(\epsilon_{\mathrm{i}},\epsilon_{\mathrm{h}}, R/a),
\label{eq:scale}
\end{equation}
where $f_{\perp}$ and $f_{\,\|}$ are functions of
$\epsilon_{\mathrm{i}}$, $\epsilon_{\mathrm{h}}$ and the ratio
$R/a$.

It is of interest to compare these plots with the same forces as
calculated in the dipole-dipole approximation. For two parallel
dipoles ${\bf p}_1$ and ${\bf p}_2$, located at the origin and at
${\bf R} = R\,{\bf \hat{z}}$, the $z$ component of the force acting
on the sphere at ${\bf R}$ has the well-known form
\begin{equation}
F_{12}^{\mathrm{dip},\|}(R) = 3\,\frac{\,p_1p_2}{R^4}\left[1 -
3\,({\bf \hat{p}}_1\cdot{\bf \hat{z}})^2\right],
\end{equation}
where $p_1$ and $p_2$ are the magnitudes of the two dipole moments,
and ${\bf \hat{p}}_1$ is a unit vector parallel to ${\bf p}_1$ (or
${\bf p}_2$). For the present case, if the spheres are well
separated and have equal radii, the dipole moments can be calculated
as if each is an isolated sphere in a uniform external electric
field ${\bf E}_0$:
\begin{equation}
{\bf p}_1 = {\bf p}_2 = a^3{\bf E}_0\frac{\epsilon_{\mathrm{i}} -
1}{\epsilon_{\mathrm{i}} + 2}.
\end{equation}
For the cases in which the unit vector  ${\bf \hat{E}}_0$ is
perpendicular and parallel to ${\bf \hat{z}}$, the radial component
of the force reduces to
\begin{equation}
F_{12}^{\mathrm{dip},\perp} = -\frac{1}{2}\,F_{12}^{\mathrm{dip},\|}
= 3\,a^6E_0^{\,2}\left[\frac{\epsilon_{\mathrm{i}}-1}
{\epsilon_{\mathrm{i}}+2}\right]^{2}\frac{1}{R^{\,4}}.
\label{eq:fdipole}
\end{equation}
These values of $F_{12}^{\mathrm{dip},\perp}$ and
$F_{12}^{\mathrm{dip},\|}$ shown in Fig.\ \ref{fig:eqsize} agree
very well with those calculated from eq.\ (\ref{eq:fdipole}) (taking
$\epsilon_{\mathrm{i}} = \infty$) at large separation ($R \gg a$)
but depart strongly at small separation
$(\delta\equiv R - 2\,a \ll a)$.  Just as
in the exact calculation, the radial component of the force in the
dipole-dipole limit is repulsive in the perpendicular case and
attractive in the parallel case.
However, the ratio of the two forces in the parallel and
perpendicular configurations at small separation has a magnitude
greater than $50$ for $R = 0.632\,\mathrm{cm} = 2\,a +
0.002\,\mathrm{cm}$,
which is much larger than the factor of $2$ expected
from the dipole-dipole approximation.  Further examples of this
ratio are given in Table I for various separations.

\begin{figure}
  \hfill
  \begin{minipage}[t]{.47\textwidth}
    \begin{center}
      \epsfig{file=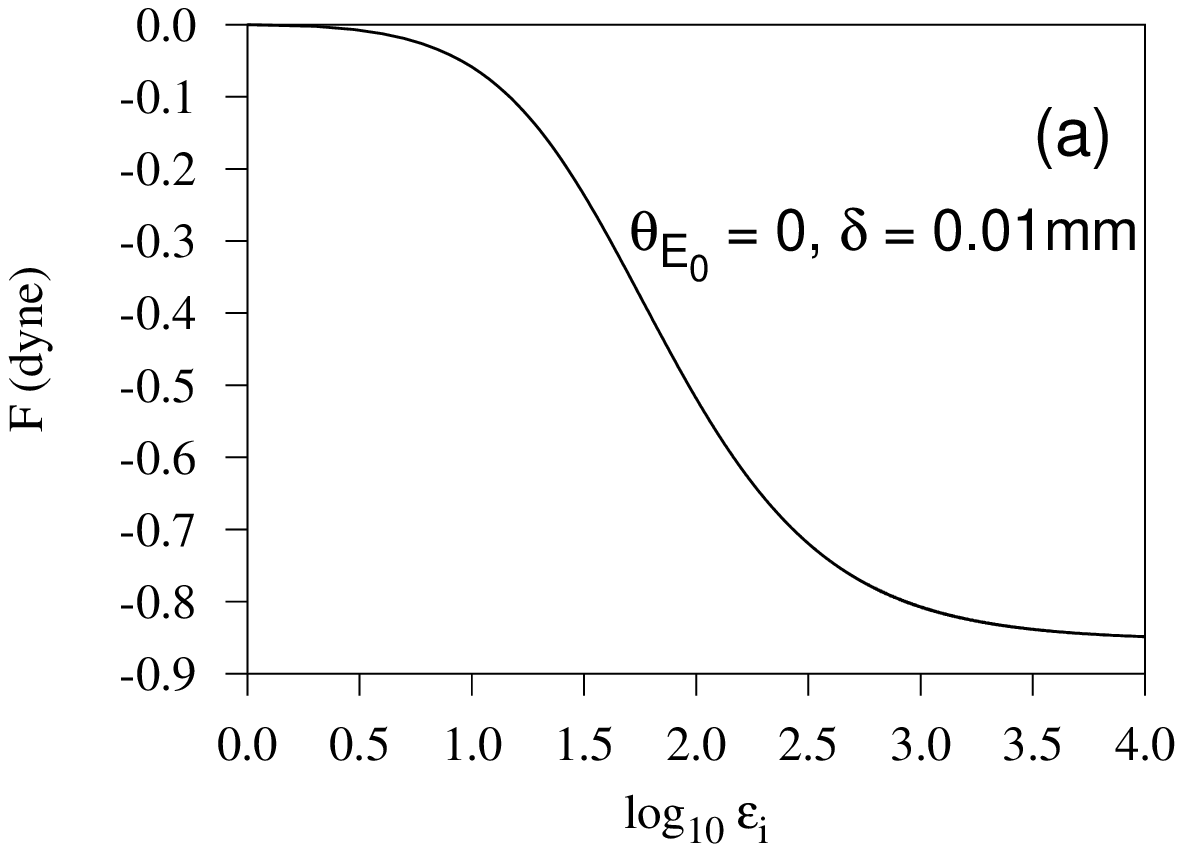, scale=0.6}
    \end{center}
  \end{minipage}
  \hfill
  \begin{minipage}[t]{.47\textwidth}
    \begin{center}
      \epsfig{file=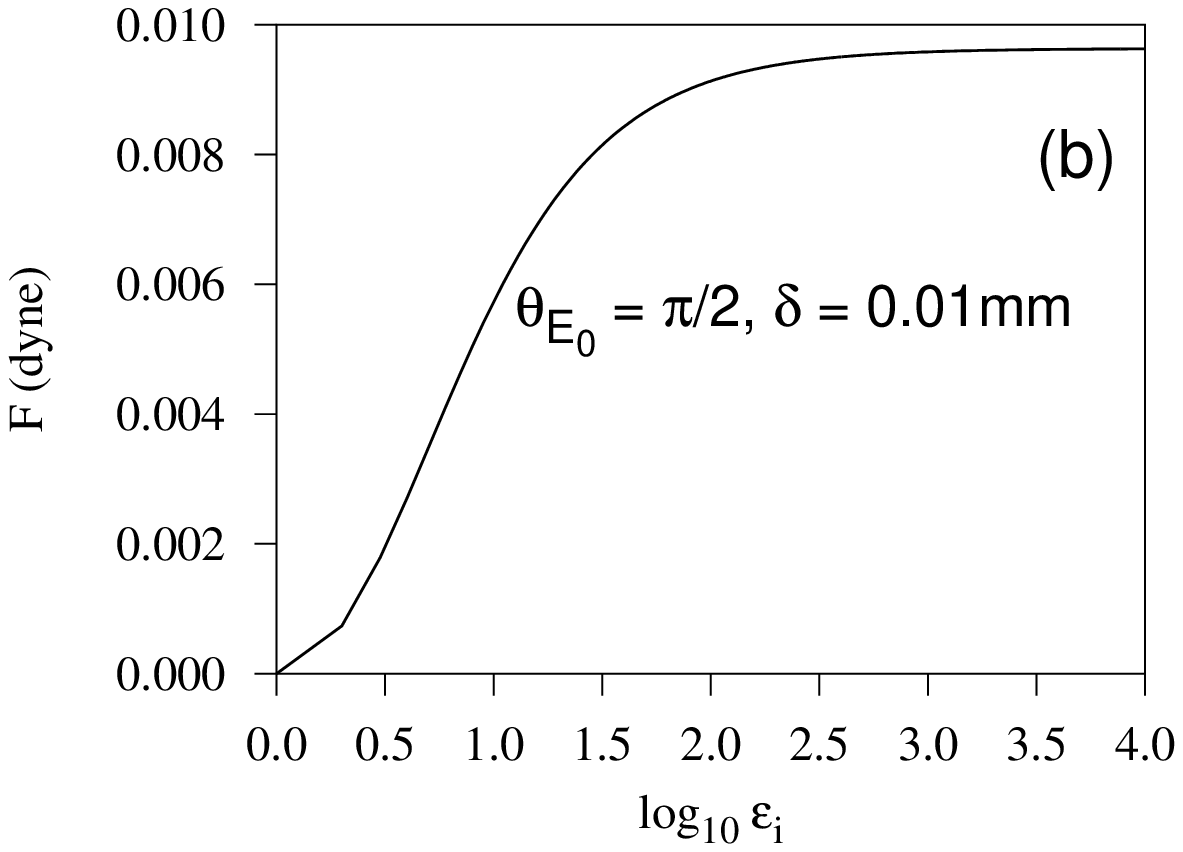, scale=0.6}
    \end{center}
  \end{minipage}
  \hfill
\caption{\label{fig:seich} \footnotesize{The radial component of
the force at zero frequency between two identical spheres of
dielectric constant $\epsilon_{\mathrm{i}}$, radius
$a=3.15\,\mathrm{mm}$, in a host of dielectric constant
$\epsilon_{\mathrm{h}} = 1$, at an intersphere spacing
(surface-to-surface separation) of $0.01\,\mathrm{mm}$, plotted as
a function of $\epsilon_{\mathrm{i}}$, for (a) electric field
parallel to the axis between spheres, and (b) field perpendicular
to that axis. We assume an electric field of strength
$25.2\,\mathrm{V\!/mm}$. Negative and positive forces denote
attractive and repulsive forces, respectively.}}
\end{figure}

\begin{figure}
  \hfill
  \begin{minipage}[t]{.47\textwidth}
    \begin{center}
      \epsfig{file=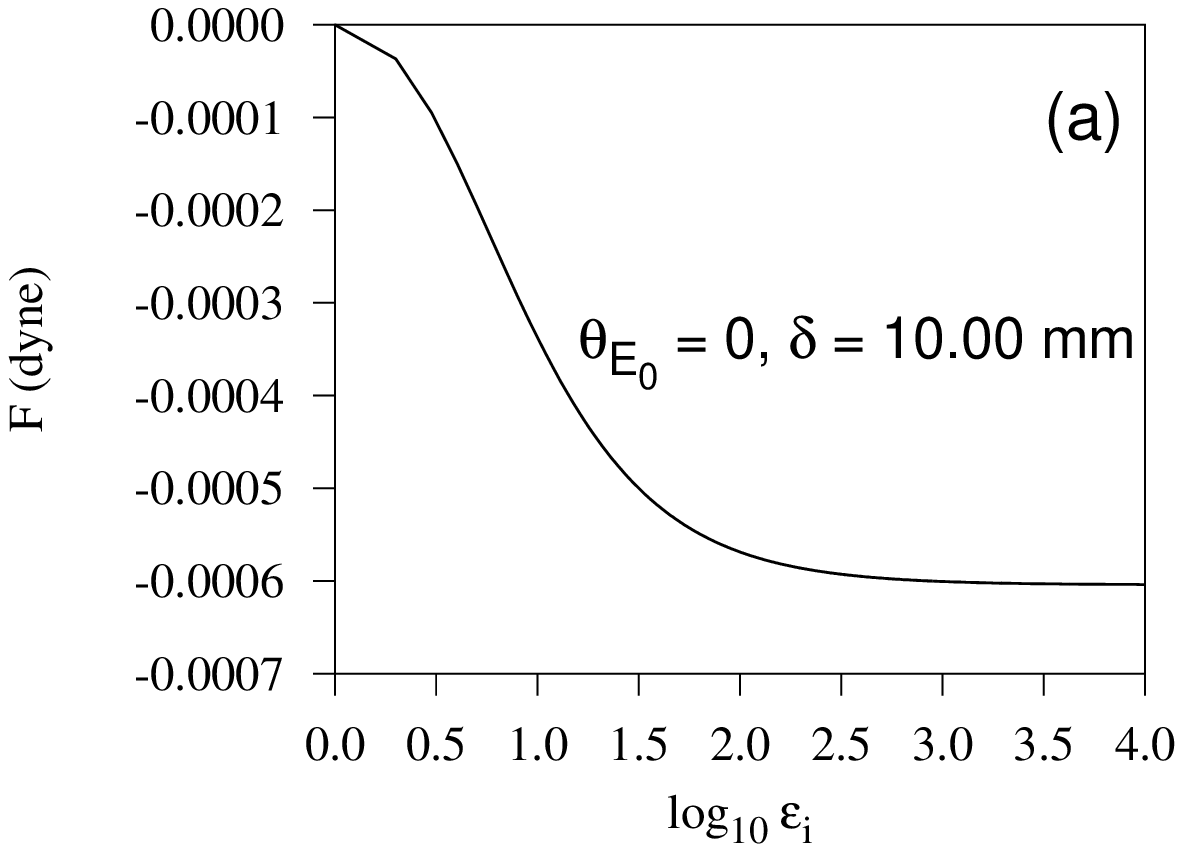, scale=0.6}
    \end{center}
  \end{minipage}
  \hfill
  \begin{minipage}[t]{.47\textwidth}
    \begin{center}
      \epsfig{file=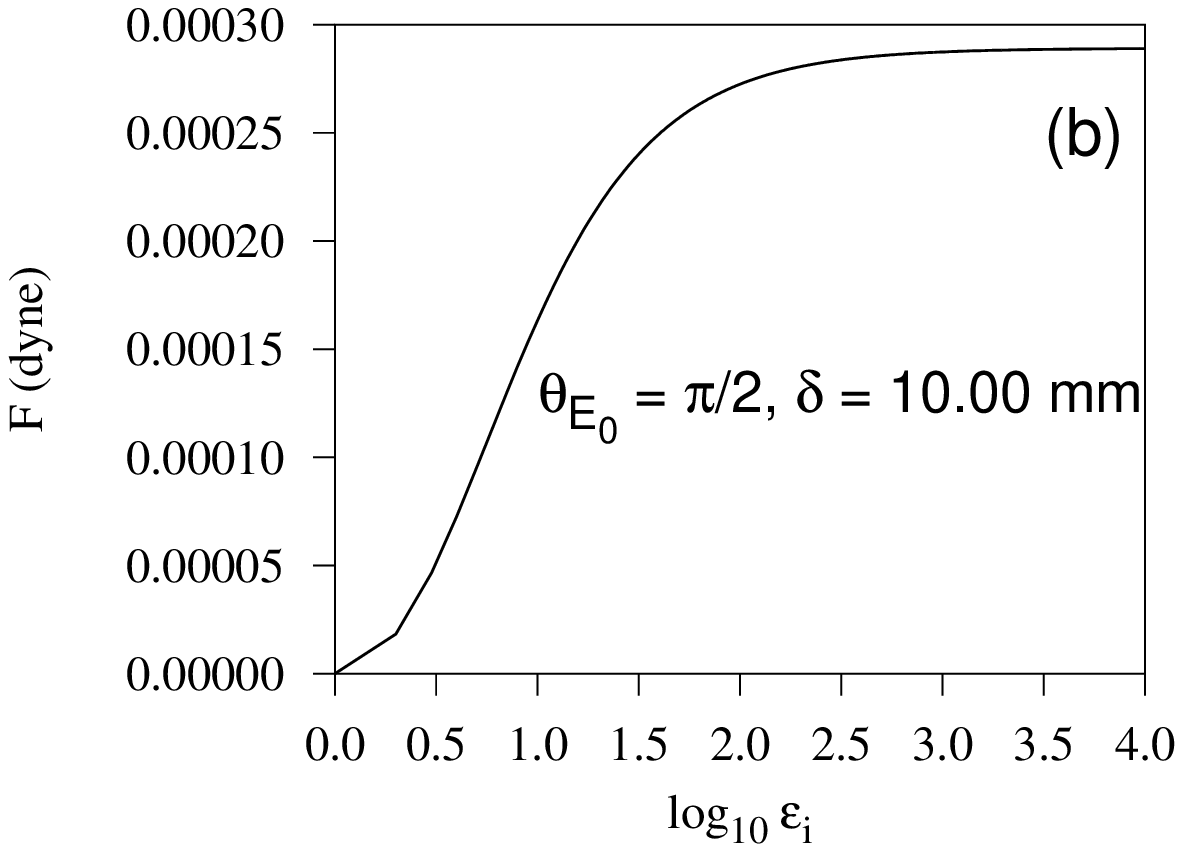, scale=0.6}
    \end{center}
  \end{minipage}
  \hfill
\caption{\label{fig:leich} \footnotesize{Same as Fig.\
\ref{fig:seich}, but for an intersphere spacing $\delta =
10.00\,\mathrm{mm}$.}}
\end{figure}

In Figs.\ \ref{fig:seich} and \ref{fig:leich}, we test the effect
of different inclusion dielectric constants, by calculating the
force between two identical spheres, each of radius $a$ and
dielectric constant $\epsilon_{\mathrm{i}}$, in a host of
dielectric constant $\epsilon_{\mathrm{h}} = 1$.  We plot the
radial component of this force, for both the parallel and
perpendicular configurations, as a function of
$\epsilon_{\mathrm{i}}$, for two different separations between the
spheres: $R = 2\,a + 0.01\,\mathrm{mm}$ and $R = 2\,a +
10.00\,\mathrm{mm}$, where we again use $a = 3.15\,\mathrm{mm}$.
In the second case, the forces are very close to the dipole-dipole
predictions.
In the first case, the forces exhibit a large departure from the
predictions of the dipole-dipole interaction, and this departure
becomes greater as $\epsilon_{\mathrm{i}}$ deviates more and more
from unity.

Next, we consider an example in which the dielectric functions of
both the inclusion and the host depend on frequency. Specifically,
we choose
\begin{equation}
\epsilon_{\mathrm{i}} = \epsilon_{\mathrm{i0}} + i\,\frac{4\pi
\sigma_{\mathrm{i}}}{\omega}, \label{eq:dfincl}
\end{equation}
and
\begin{equation}
\epsilon_{\mathrm{h}} = \epsilon_{\mathrm{h0}} + i\,\frac{4\pi
\sigma_{\mathrm{h}}}{\omega}, \label{eq:dfhost}
\end{equation}
where $i$ is the imaginary unit, $\epsilon_{\mathrm{i0}}$ and
$\epsilon_{\mathrm{h0}}$ are the dielectric constants of the
inclusion and the host, and $\sigma_{\mathrm{i}}$ and
$\sigma_{\mathrm{h}}$ are their conductivities, assumed
frequency-independent.  The {\em time-averaged} forces are now
calculated  from the generalization of eq.\ (\ref{eq:force2}) to
finite frequencies, namely, eq.\ (\ref{eq:fcomplex}).

We have chosen to use parameters given by Ref.\ \cite{zhiyong}, in a
recent experimental study.
These are listed in Table II.  However, as discussed further below,
it is possible that the experimentally measured forces include
effects beyond the purely electrostatic interactions included in our
model (such as spatially dependent conductivities of the host
fluid).  Therefore, our numerical results should again be considered
as model calculations, not necessarily applicable to the specific
experiments of Ref.\ \cite{zhiyong}.  In all cases, we assume that
the two spherical inclusions are identical, with a dielectric
constant and conductivity characteristic of $\mathrm{SrTiO_{3}}$.
For the host fluid, we have considered the various materials used in
the measurements of Ref.\ \cite{zhiyong}. (In practice, the nonzero
conductivity of SrTiO$_3$ has negligible effect on force; we have
checked this by recalculating the forces with the conductivity set
equal to zero, and obtained the same results.)

In Fig.\ \ref{fig:sc} (a) and (b), we show the radial component of
the calculated time-averaged force on a sphere of
$\mathrm{SrTiO_{3}}$ at $\mathbf{R}$ in the parallel and
perpendicular geometries, for the host materials of silicone oil
and castor oil.  In both cases, we assume spheres of radius $a =
3.15\,\mathrm{mm}$, intersphere spacing $\delta=0.01\,\mathrm{mm}$
and applied electric field $E_{0}=25.2\,\mathrm{V\!/mm}$, as in
Ref.\ \cite{zhiyong}. The magnitude of force decreases with
increasing frequency, but rapidly converges to a constant value in
both cases.   The sign of the force is negative in (a),
corresponding to an attractive force, and positive (repulsive) in
(b).

If these were strictly dipole-dipole forces, the time-averaged
force on the sphere at ${\bf R}$ would be given by the
generalization of eq.\ (\ref{eq:fdipole}) to complex dielectric
functions and $\epsilon_{\mathrm{h}} \neq 1$, namely
\begin{equation}
F_{\mathrm{av},12}^{\mathrm{dip},\perp} =
-\frac{1}{2}\,F_{\mathrm{av},12}^{\mathrm{dip},\|} =
\frac{3}{2R^{\,4}}\:a^6E_0^{\,2}\:\mathrm{Re}\!
\left[\left(\frac{\epsilon_{\mathrm{i}}-\epsilon_{\mathrm{h}}}
{\epsilon_{\mathrm{i}}+2\,\epsilon_{\mathrm{h}}}\right)^{\!\!2}
\epsilon_{\mathrm{h}}\right]. \label{eq:force1a}
\end{equation}
Thus, in particular, the magnitude of the force in the parallel
case would be twice as large as that in the perpendicular case, as
in our previous examples.  However, in  Fig.\ \ref{fig:sc}, this
force ratio is about $50$.  This difference occurs, as in Fig.\
\ref{fig:eqsize}, because of the very small separation
($\delta=0.01\,\mathrm{mm}$), which corresponds to a very
short-ranged interaction. In the long range limit ($R \gg a$), our
calculated magnitude ratio agrees well with the dipole-dipole
prediction, as discussed further below.  This short-distance
deviation from dipole-dipole forces is similar to that seen in
Figs.\ \ref{fig:eqsize}--\ref{fig:leich}.

Fig.\ \ref{fig:sc} also shows that there is a substantial
difference between the forces for silicone oil and castor oil
hosts.  This difference is due almost entirely to the difference
in the static dielectric constants of these two hosts: the effect
of the finite conductivity disappears by about $10\,\mathrm{Hz}$
in both cases, whereas the difference between the forces persists
to much higher frequencies.

The calculated time-averaged force between spheres of SrTiO$_3$ in
a silicone oil host is plotted versus separation in Fig.\
\ref{fig:distance} at a frequency of $50\,\mathrm{Hz}$.  In order
to see the effects of a finite host conductivity, we include this
conductivity in Figs.\ \ref{fig:distance} (a) and
\ref{fig:distance} (b) but not in \ref{fig:distance} (c) or
\ref{fig:distance} (d).  We also set the conductivity of the
sphere equal to zero in (c) and (d).
Clearly, the host conductivity has very little influence on the
forces at this frequency.  For comparison, we also show the forces
as calculated in the dipole-dipole approximation.  As can be seen,
there is very little difference between the two except for $R
<\sim 1.5\,\mathrm{cm}$.
Even at such small spacings, the deviation from the dipole-dipole
force is much larger for the parallel than the perpendicular
configuration. At a spacing of $0.01\,\mathrm{mm}$, the calculated
ratio of force magnitudes in the parallel and perpendicular
configurations exceeds a factor of $100$.

At sufficiently high host conductivity, our model predicts that the
force between spheres changes sign as a function of frequency. This
trend is shown in Fig.\ \ref{fig:ebesms} for a separation of
$\delta=0.01\,\mathrm{mm}$ between spheres. The host materials used
here are ethyl benzoate, ethyl salicylate, and methyl salicylate,
all of which have much greater conductivities than silicone oil. The
sign change is due mainly to the greater conductivities, not the
differences in static dielectric constants.  To check this point, we
recalculated the points of Fig.\ \ref{fig:ebesms} assuming the same
value of the real part of the dielectric constant for all three host
materials; we found that the time-averaged forces changed sign at
the same frequencies as in Fig.\ \ref{fig:ebesms}. Mathematically,
the origin of the sign change is, of course, the dependence of the
variable $s$ in eqs.\ (\ref{eq:s}) and (\ref{eq:fcomplex}) on the
host conductivity.

The time-averaged force for this separation ranges from about
$+1.5$ to $-1.5\,\mathrm{dynes}$ for the parallel case, depending
on the frequency, and from about $+0.5$ to $-3.0\,\mathrm{dynes}$
for the perpendicular case.
At high frequencies, the force approaches $-1.0\,\mathrm{dyne}$
for the parallel case, whatever the host fluid is, and approaches
a much smaller magnitude in the perpendicular case. The ratio of
these forces differs greatly from the predictions of the
dipole-dipole interaction, as expected for such a small
separation. At very low frequencies, however, the force ratio
appears to approach the dipole-dipole prediction.

Fig.\ \ref{fig:sn2} shows the frequency dependence of the
time-averaged force between two spheres of SrTiO$_3$ for silicone
oil and $N_2$ hosts.   Both the spacings $\delta$ between the two
spheres and the electric field $E_0$ are larger than those for Fig.\
\ref{fig:sc}; they are given in the legends of each Figure. We chose
these values for the parameters because they are used in the
measurements of Ref.\ \cite{zhiyong}.
Evidently, the force between the two spheres is stronger when the
two spheres are immersed in a liquid host than in a gas, all the
other parameters of the forces being held constant.  This behavior
can be understood even in the dipole-dipole limit: it is due to
the dependence of the force on $\epsilon_{\mathrm{h}}$ as in eq.\
(\ref{eq:force1a}). Also, the low-frequency forces in Fig.\
\ref{fig:sn2} (a) and (b) and especially (c) and (d) depend more
weakly on frequency than those in Fig.\ \ref{fig:sc}. Another
point is that, even though the intersphere spacing $\delta$ has
been increased to $0.10$ and $0.30\,\mathrm{mm}$ in these
calculations, the calculated forces are still far from the
dipole-dipole limit. Specifically, the ratio of the force
magnitudes in the parallel and perpendicular geometries greatly
exceeds the factor of two expected in the dipole-dipole limit.
However, this ratio is smaller than that of Fig.\ \ref{fig:sc},
presumably because the intersphere separations are larger than in
that Figure.

\section{\label{sec:level4}Discussion}

The present work permits calculation of electrical forces in ER
fluids in a concise closed form, which permits inclusion of all
multipoles and all many-body forces in a simple way.  In our
approach, the forces do not need to be calculated as numerical
derivatives; instead, we give explicit analytical expressions for
these derivatives, in terms of a pole spectrum which characterizes
the microgeometry of the material. The explicit form for the
derivatives is somewhat reminiscent of the Hellmann-Feynman
description of quantum-mechanical forces in electronic structure
theory, but differs from it in the important respects.

One striking feature of the present formalism is that it allows
for the calculation of frequency-dependent forces in a simple
closed form.
Although such forces have been discussed in previous
work\cite{davis3,tang1,tang2,ma}, the present approach is relatively
simple and more general, and places both zero and finite frequency
forces within the same formalism.
In our numerical work, we find that these forces can even change
sign as a function of frequency. Such frequency-dependence is, of
course, also present in the long-range (dipole-dipole) limit
treated by others in the previous work, but it is even more
apparent in the present study.

Although in the present work calculations have been carried out
explicitly for two-body interaction, they can readily be extended
to three-body (or multi-body) forces.  The general equation
(\ref{eq:force2}) or
(\ref{eq:fcomplex}) can be used to find the force on a sphere, no
matter how many particles are contained in the suspension. Indeed,
such multi-body forces are very likely to play important roles in
dense suspensions, where they could possibly lead to
''bond-angle-dependent'' forces analogous to
angle dependent interatomic elastic forces in liquid and solid
semiconductors.  Likewise, the calculations could be readily
extended to more complex particles (e.\ g., hollow spherical
shells), and to non-spherical particles, provided that the
requisite pole spectra and matrix elements can be calculated.
Also, although we have restricted our calculations in this paper
to the radial component of the interparticle forces, other
components can be straightforwardly computed.  Finally, the
present formalism can be immediately extended to the important
case of magnetorheological fluids.  For such fluids, eq.\
(\ref{eq:force2}) or
(\ref{eq:fcomplex}) for the force would continue to be valid,
provided that $\epsilon_{\mathrm{i}}$ and $\epsilon_{\mathrm{h}}$
are replaced by $\mu_{\mathrm{i}}$ and $\mu_{\mathrm{h}}$.

Our calculated frequency-dependent forces, obtained using parameters
quoted for SrTiO$_3$ spheres in a conducting host, may appear to
disagree with those obtained in Ref.\ \cite{zhiyong} at close
spacing. One possible explanation for this discrepancy is that the
host fluid does not
exhibit its usual bulk conductivity when two highly
polarizable spheres are placed in it in close proximity. Instead,
there could well be non-linear screening effects of the
Debye-H\"{u}ckel type\cite{debye}, which would mean that the picture
of a two-component composite is simply not appropriate in this
regime. In support of this hypothesis, we note that the reported
experimental forces are still frequency-dependent at high
frequencies, while the complex dielectric functions of both host and
sphere should be nearly frequency-independent in this regime,
leading to a frequency-independent force in this range.


The present method could readily be combined with standard
molecular dynamics approaches to compute {\em dynamical}
properties of electrorheological (or magnetorheological) fluids.
Specifically, one could carry out molecular dynamics (MD)
calculations, following the approach of several
authors\cite{klingenberg, bonnecaze,klingenberg2,hass,tao4}. In
such approaches, the force on a given sphere is typically
expressed as the sum of a hard-sphere repulsion, a viscous force,
and an electrostatic force.  The first two of these forces would
be the same as in the previous MD studies, but the third would be
calculated using the present method, rather than the dipole-dipole
force generally used in most other MD studies. It would be of
great interest to see how such quantities as viscous relaxation
time would be affected by using our forces in these calculations.
In addition to such calculations, one could study minimum-energy
configurations of dielectric suspensions in an applied electric
field, based on the forces calculated using the methods outlined
here.  Many such studies can already be found in the literature
(see, e.\ g., Ref.\ \cite{davis5} or \cite{chaikin}). It would be
of interest to extend the present approach to calculating
minimum-energy configurations including non-dipolar forces, as
outlined in the present work.

\section{\label{sec:level5}Acknowledgments}

This work was supported by NSF Grants DMR01-04987 and DMR04-13395
(KK and DS). The work of DJB and XL was supported, in part, by
grants from the US-Israel Binational Science Foundation and the
Israel Science Foundation. The work of XL was also supported by a
grant from the Sackler Institute of Solid State Physics of Tel
Aviv University. All the calculations were carried out on the P4
Cluster at the Ohio Supercomputer Center, with the help of a grant
of time.

\newpage

\begin{center}
TABLES
\end{center}

\vspace{0.1in}
\begin{tabular}{|c|c|c|}
\hline \multicolumn{1}{|c}{$R$(cm)} & \multicolumn{1}{|c|}{Force
ratio} &
\multicolumn{1}{c|}{$(R-2\,a)/(2\,a)$} \\
\hline \hline
0.630 & 602.3  &  0.0000 \\
0.631 &  88.5  &  0.0016 \\
0.632 &  52.5  &  0.0032 \\
0.633 &  38.8  &  0.0048 \\
0.634 &  31.5  &  0.0063 \\
0.635 &  26.8  &  0.0079 \\
0.636 &  23.5  &  0.0095 \\
0.637 &  21.1  &  0.0111 \\
0.638 &  19.2  &  0.0127 \\
0.639 &  17.7  &  0.0143 \\
0.640 &  16.5  &  0.0159 \\
\hline
\end{tabular}

\vspace{0.2in}

\noindent {\footnotesize TABLE I.  The ratios of the magnitudes of
the forces between two identical spheres in the parallel and
perpendicular configurations, calculated at several small
separations and assuming $\epsilon_{\mathrm{i}} = 10^5$,
$\epsilon_{\mathrm{h}} = 1$, $a=3.15\,\mathrm{mm}$,
$E_{0}=25.2\,\mathrm{V\!/mm}$, and $\omega=0$.  The force is
attractive in the parallel configuration, repulsive in the
perpendicular configuration.}

\vspace{0.5in}

\begin{tabular}{|l|c|c|}
\hline \multicolumn{1}{|c}{Material} &
\multicolumn{1}{|c|}{Dielectric constant} &
\multicolumn{1}{c|}{Conductivity} \\
\hline \hline
SrTiO$_3$  &  249.0  &  2.0 $\times$ 10$^{-8}$ \\
Silicone oil &  2.54 &  1.0 $\times$ 10$^{-13}$\\
Castor oil & 4.20 & 1.0 $\times$ 10$^{-13}$ \\
Ethyl benzoate & 5.45 & 5.0 $\times$ 10$^{-8}$ \\
Ethyl salicylate & 8.65 & 1.0 $\times$ 10$^{-7}$ \\
Methyl salicylate & 9.46 & 6.0 $\times$ 10$^{-7}$ \\
$N_2$ gas & 1.00058 &  0 \\
\hline
\end{tabular}

\vspace{0.2in}

\begin{figure}
  \hfill
  \begin{minipage}[t]{.47\textwidth}
    \begin{center}
      \epsfig{file=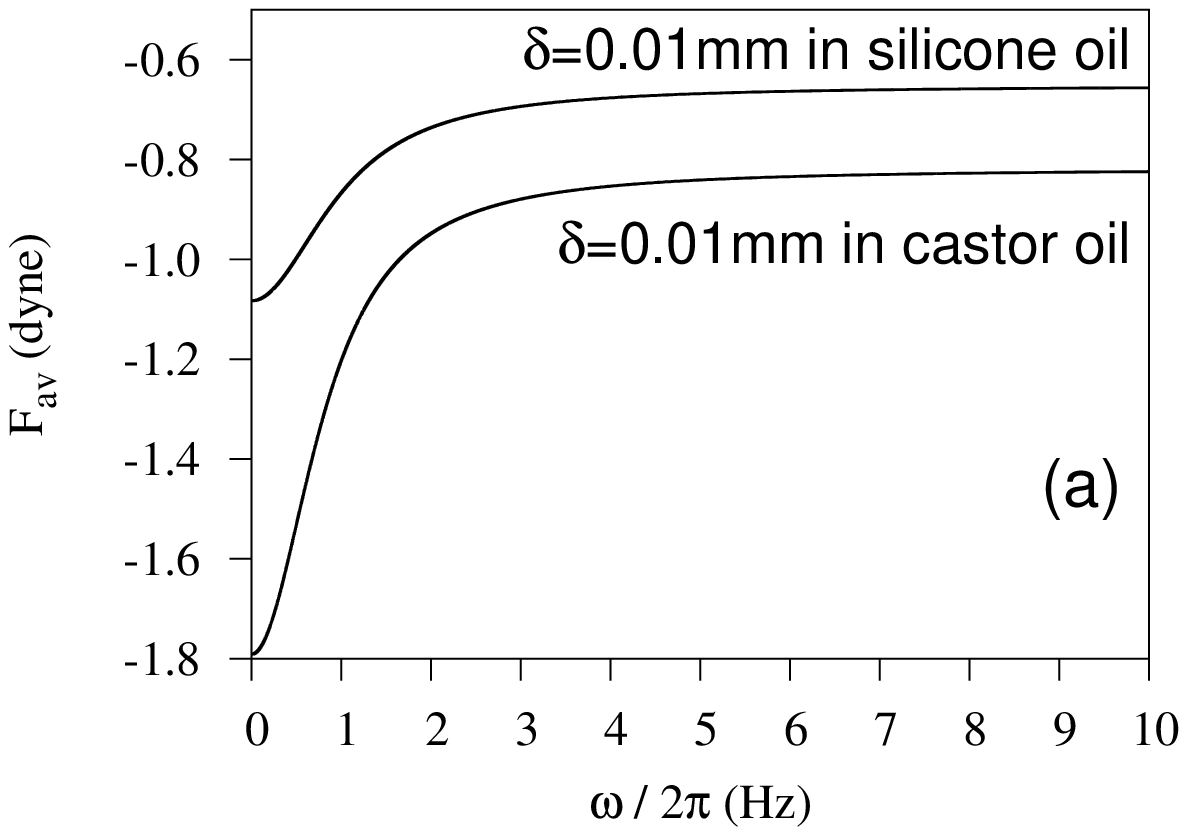, scale=0.6}
    \end{center}
  \end{minipage}
  \hfill
  \begin{minipage}[t]{.47\textwidth}
    \begin{center}
      \epsfig{file=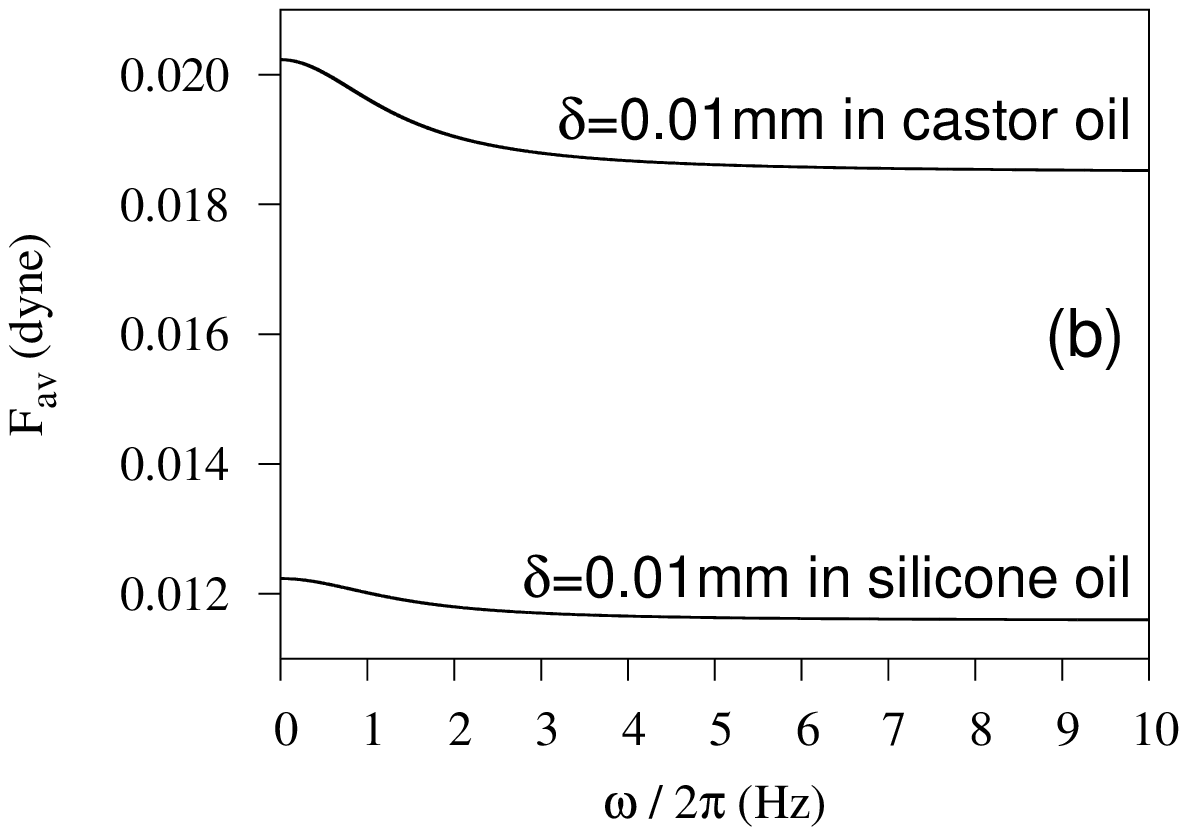, scale=0.6}
    \end{center}
  \end{minipage}
  \hfill
\caption{\label{fig:sc} \footnotesize{The radial component of the
time-averaged force between two identical spheres of
$\mathrm{SrTiO_{3}}$, plotted as a function of frequency for the
host materials of silicone oil and castor oil, respectively. For
both cases we use $\delta=0.01\,\mathrm{mm}$, $a=3.15\,\mathrm{mm}$,
and $E_{0}=25.2\,\mathrm{V\!/mm}$.  The electric field is parallel
to the line connecting the spheres in (a) and perpendicular to that
line in (b). A negative value denotes an attractive force.}}
\end{figure}

\begin{figure}
\begin{tabular}{cc}
  \hfill
  \begin{minipage}[t]{.47\textwidth}
    \begin{center}
      \epsfig{file=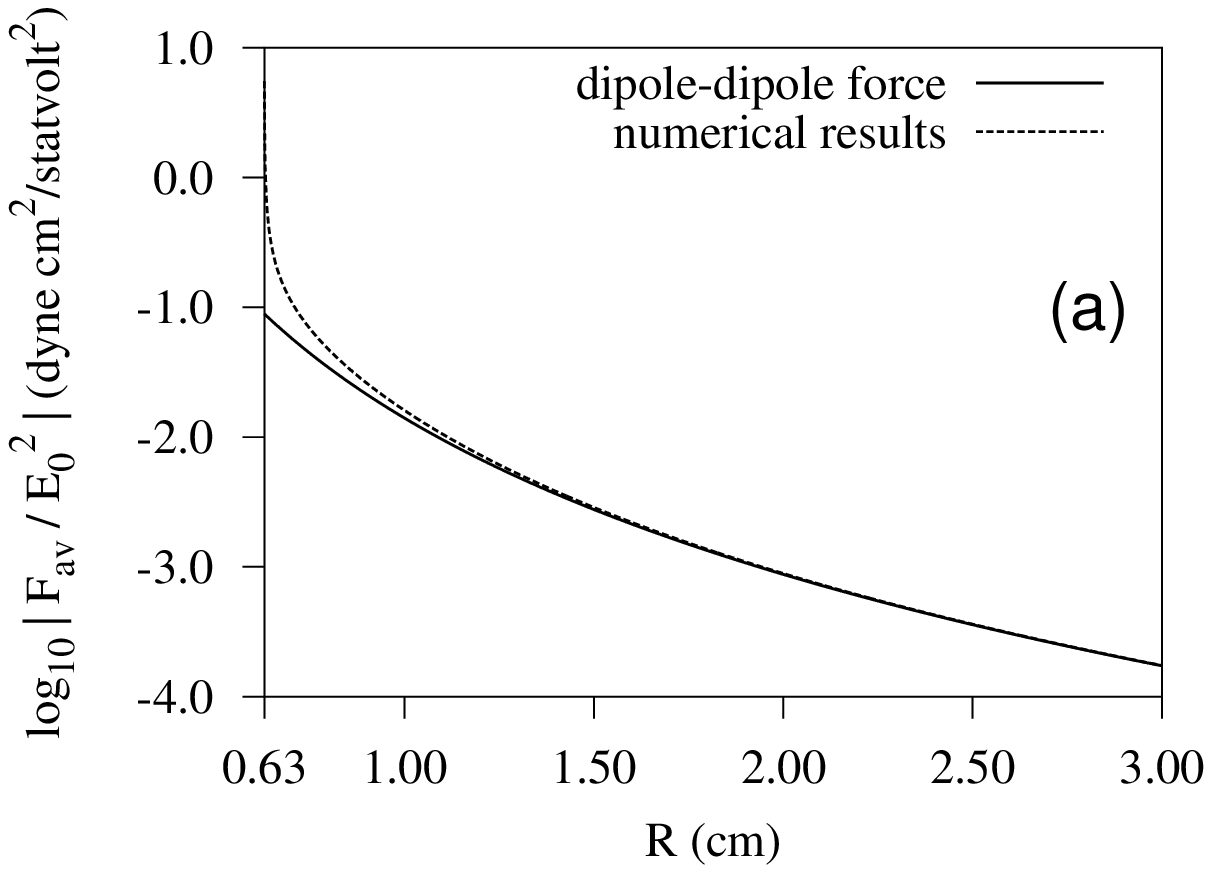, scale=0.6}
    \end{center}
  \end{minipage} &
  \hfill
  \begin{minipage}[t]{.47\textwidth}
    \begin{center}
      \epsfig{file=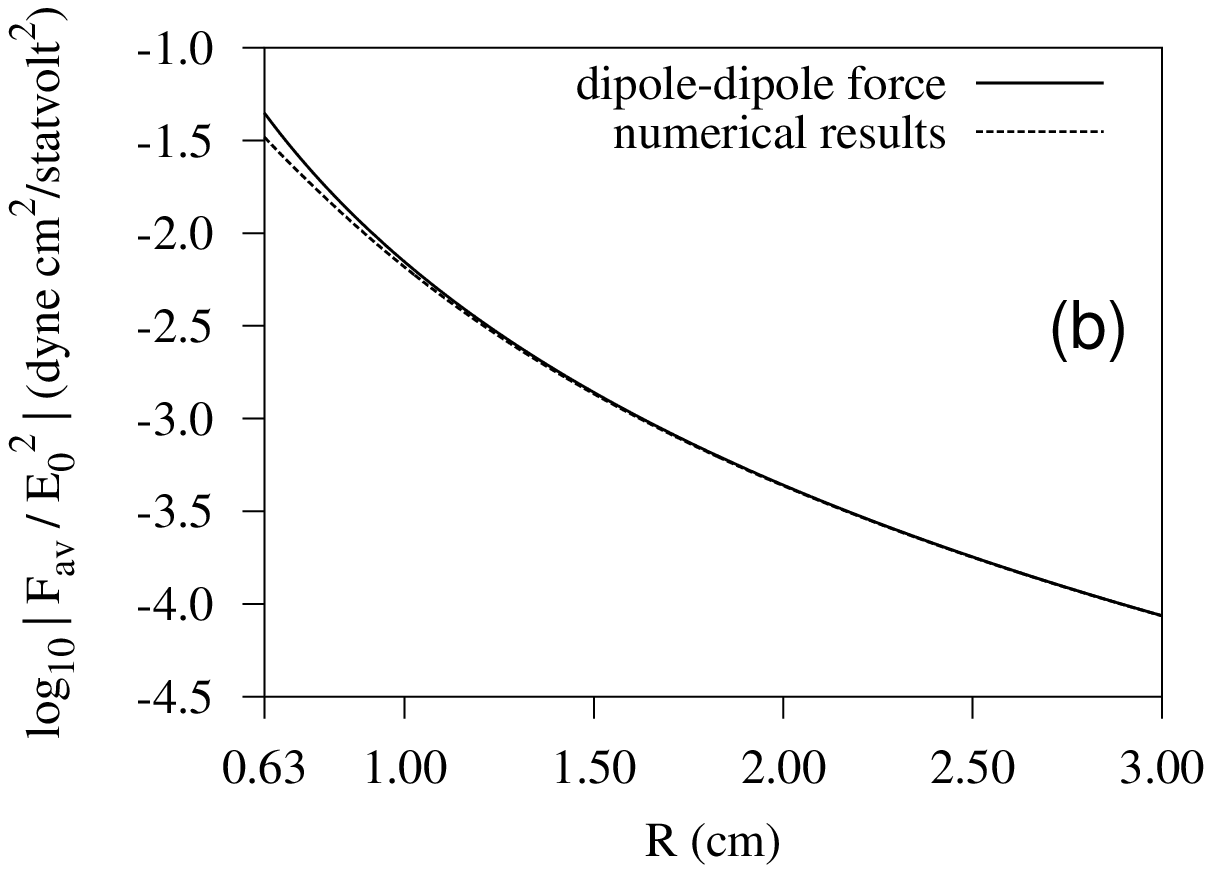, scale=0.6}
    \end{center}
  \end{minipage} \\
  \hfill
  \begin{minipage}[t]{.47\textwidth}
    \begin{center}
      \epsfig{file=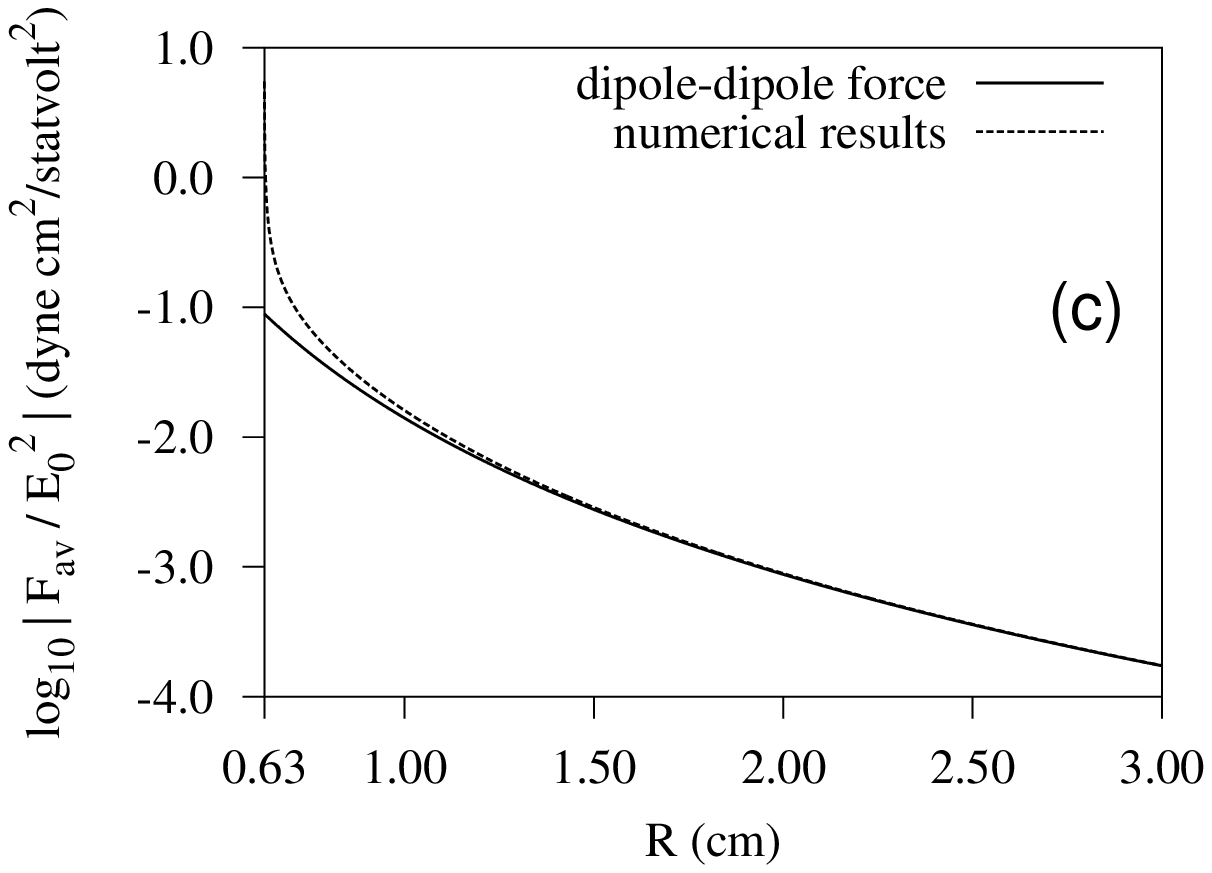, scale=0.6}
    \end{center}
  \end{minipage} &
  \hfill
  \begin{minipage}[t]{.47\textwidth}
    \begin{center}
      \epsfig{file=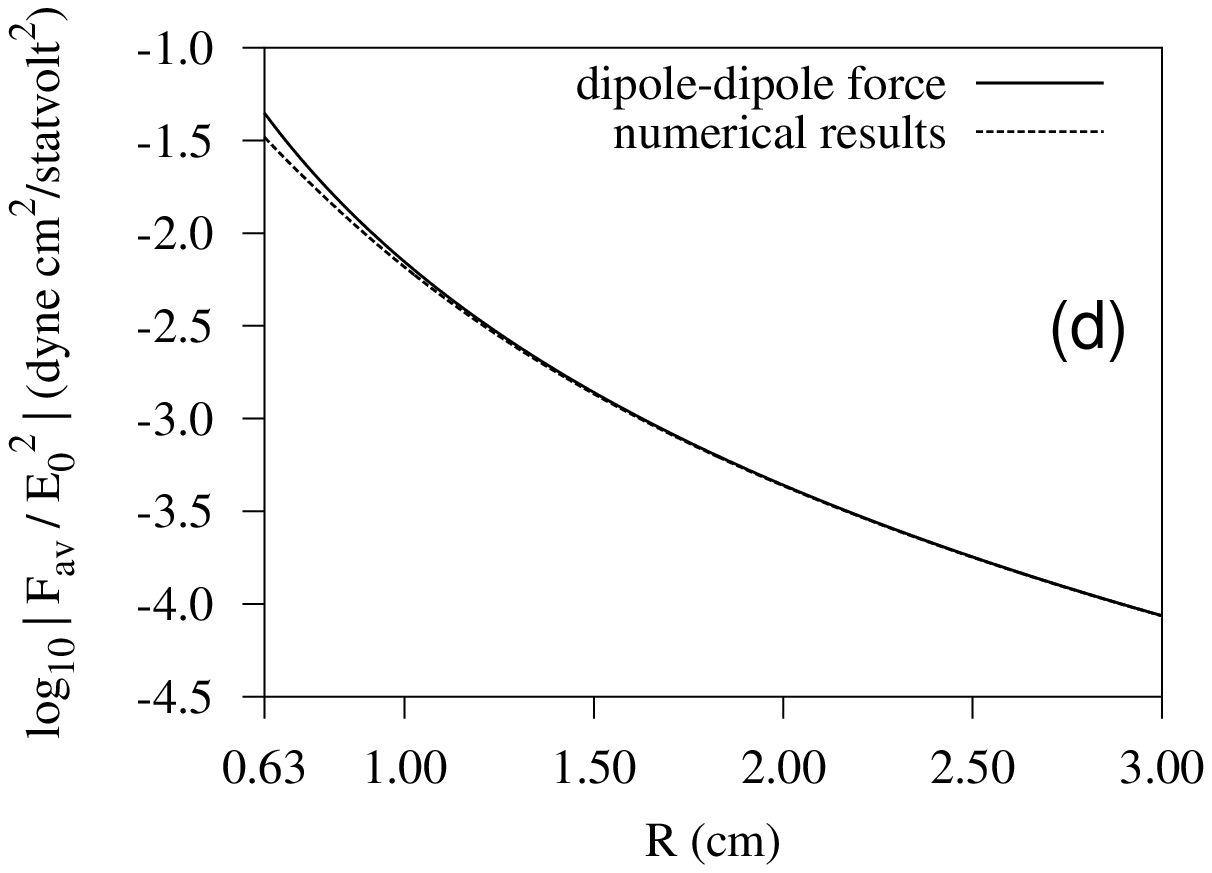, scale=0.6}
    \end{center}
  \end{minipage}
  \end{tabular}
\caption{\label{fig:distance} \footnotesize{(a) and (b) : Magnitude
of the radial component of the time-averaged force between two
identical spheres of $\mathrm{SrTiO_{3}}$, divided by $E_{0}^{\,2}$.
Also plotted is the corresponding quantity in the dipole-dipole
approximation.  Both are plotted on logarithmic scale as a function
of separation $R$ for a host material of silicone oil and a fixed
frequency of $50\,\mathrm{Hz}$. The spheres have radii
$3.15\,\mathrm{mm}$. The electric field is parallel to the line
between the two spheres in (a) and perpendicular to that line in
(b). (c) and (d) : Same as (a) and (b) except that the
conductivities of the spheres and the host are set equal to zero in
these calculations. The forces are attractive in (a) and (c),
repulsive in (b) and (d).}}
\end{figure}

\begin{figure}
  \hfill
  \begin{minipage}[t]{.47\textwidth}
    \begin{center}
      \epsfig{file=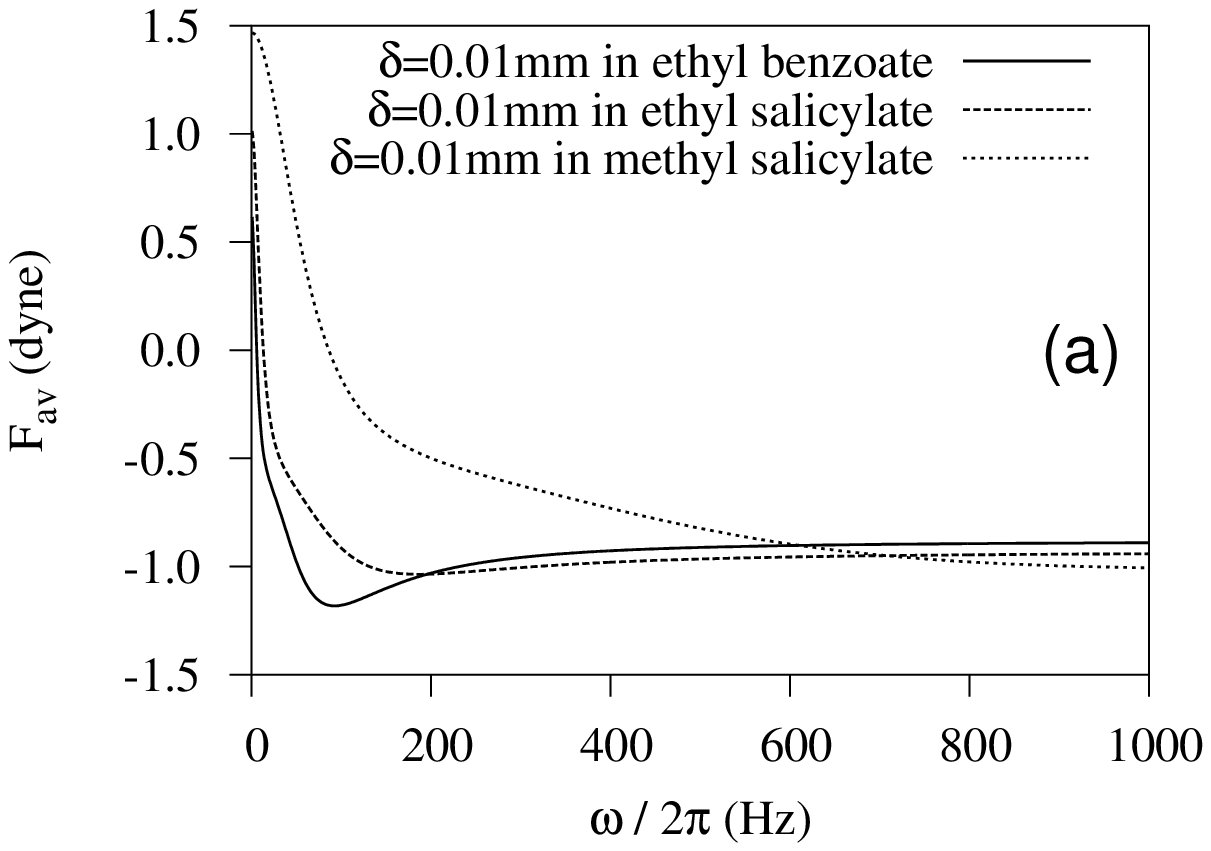, scale=0.6}
    \end{center}
  \end{minipage}
  \hfill
  \begin{minipage}[t]{.47\textwidth}
    \begin{center}
      \epsfig{file=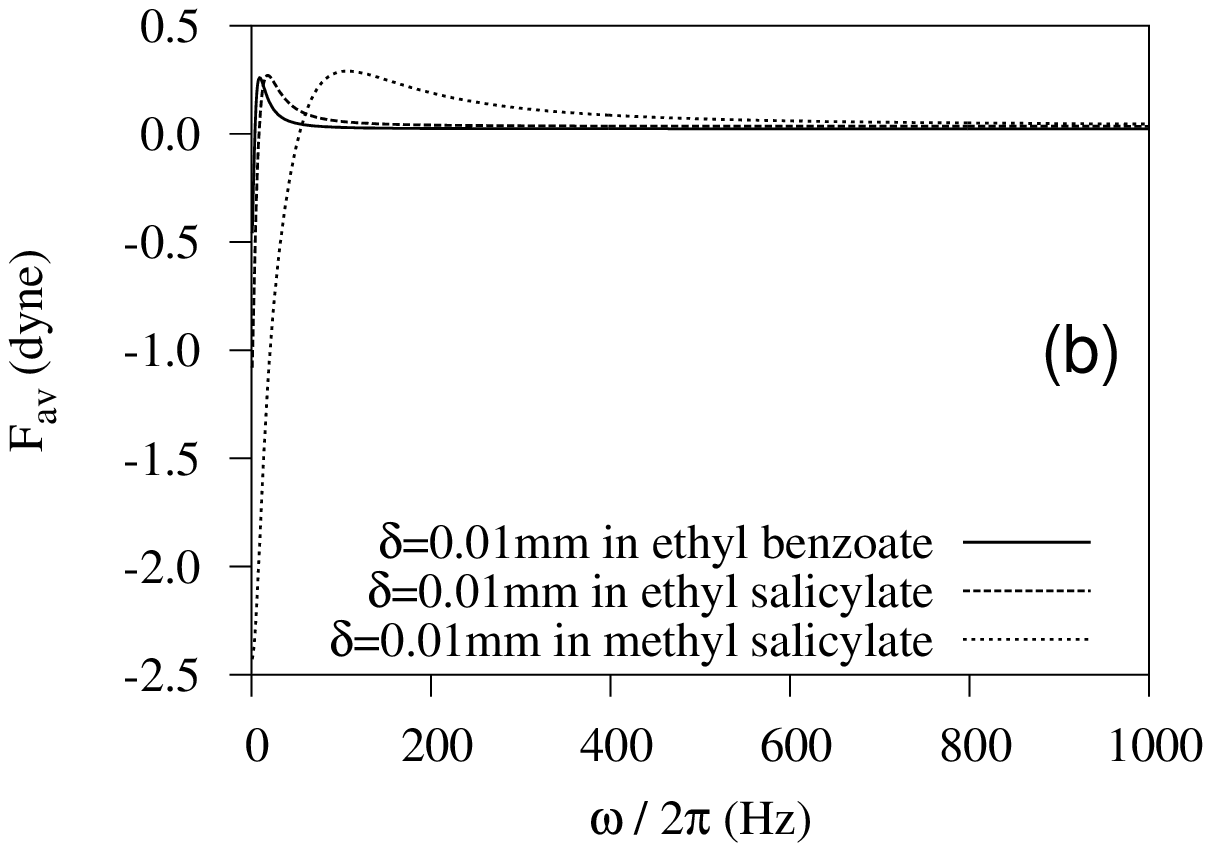, scale=0.6}
    \end{center}
  \end{minipage}
  \hfill
\caption{\label{fig:ebesms} \footnotesize{The radial component of
the time-averaged force between two identical spheres of
$\mathrm{SrTiO_{3}}$ separated by $R$, plotted as a function of
frequency for the host materials of ethyl benzoate, ethyl
salicylate, and methyl salicylate, respectively. The electric field
is parallel to the line connecting the two spheres in (a) and
perpendicular to that line in (b). In all cases,
$\delta=0.01\,\mathrm{mm}$, $a=3.15\,\mathrm{mm}$, and
$E_{0}=25.2\,\mathrm{V\!/mm}$.}}
\end{figure}

\begin{figure}
\begin{tabular}{cc}
  \hfill
  \begin{minipage}[t]{.47\textwidth}
    \begin{center}
      \epsfig{file=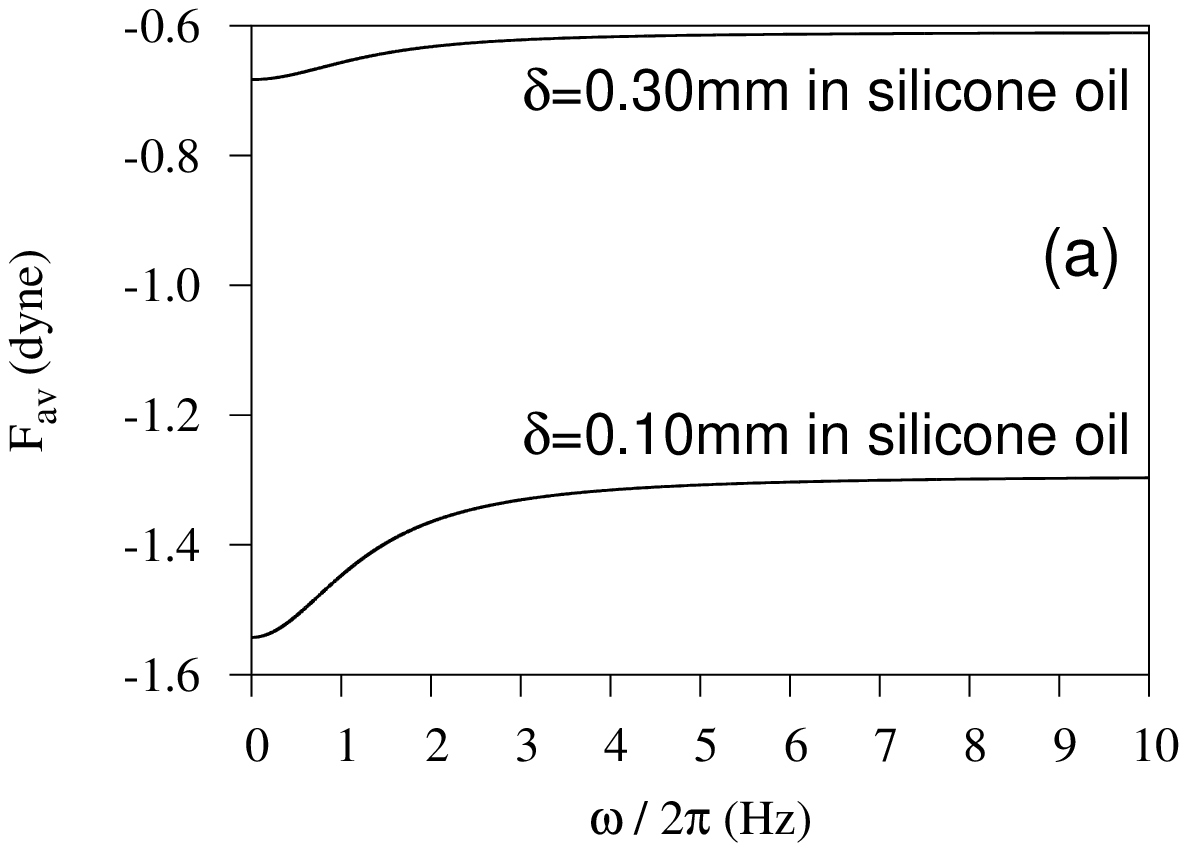, scale=0.6}
    \end{center}
  \end{minipage} &
  \hfill
  \begin{minipage}[t]{.47\textwidth}
    \begin{center}
      \epsfig{file=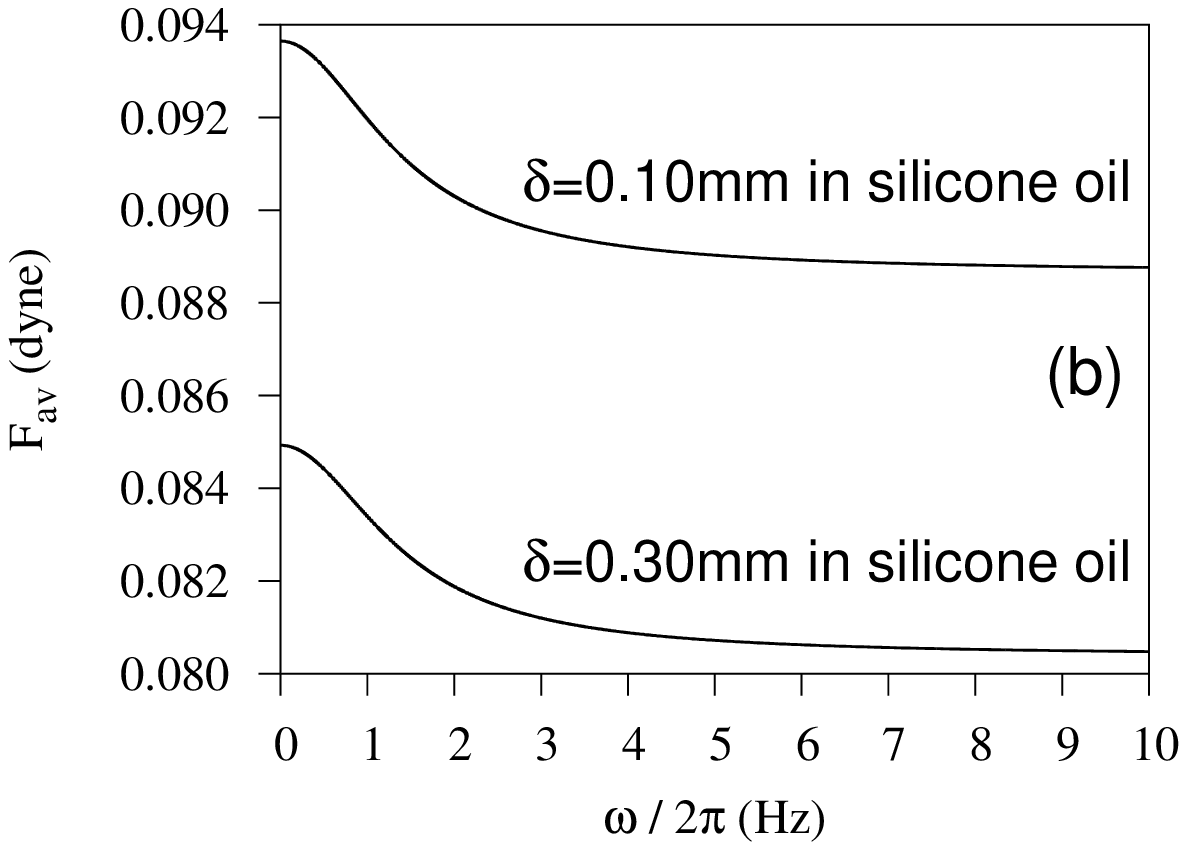, scale=0.6}
    \end{center}
  \end{minipage} \\
  \hfill
  \begin{minipage}[t]{.47\textwidth}
    \begin{center}
      \epsfig{file=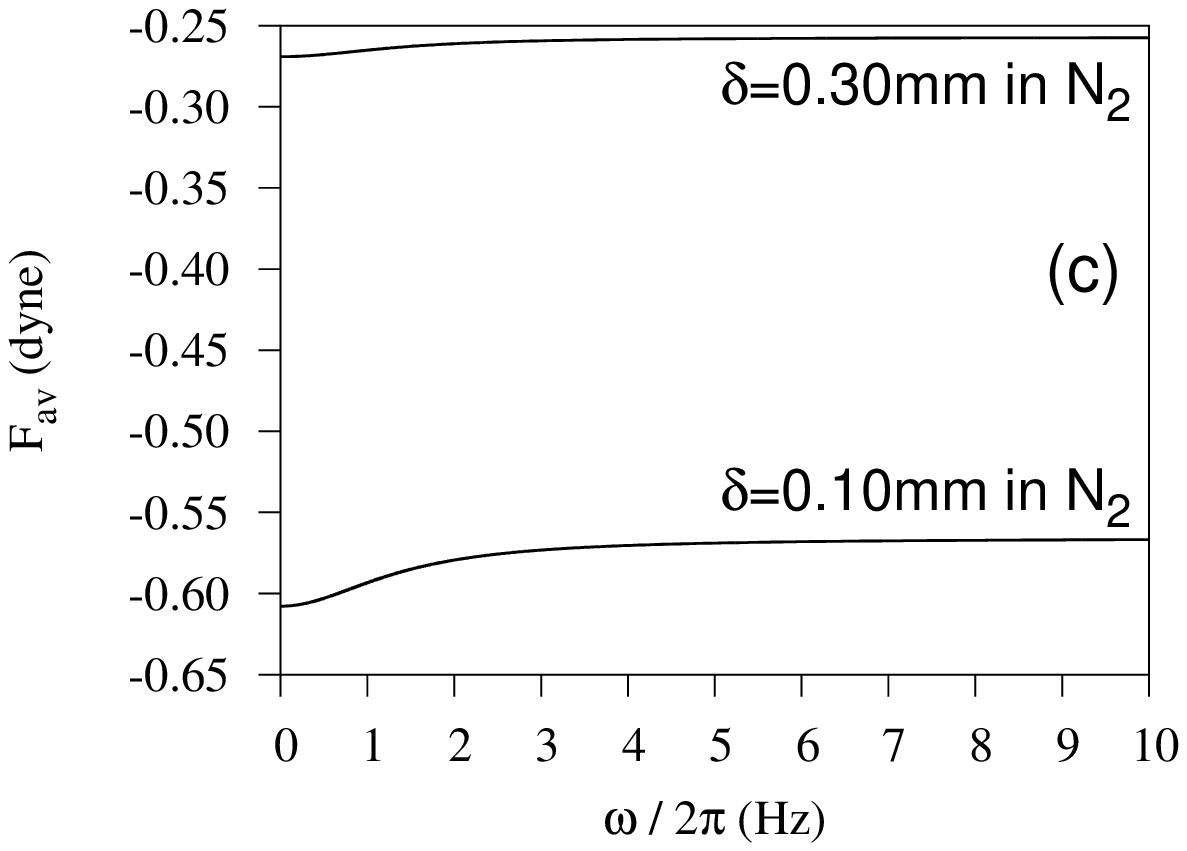, scale=0.6}
    \end{center}
  \end{minipage} &
  \hfill
  \begin{minipage}[t]{.47\textwidth}
    \begin{center}
      \epsfig{file=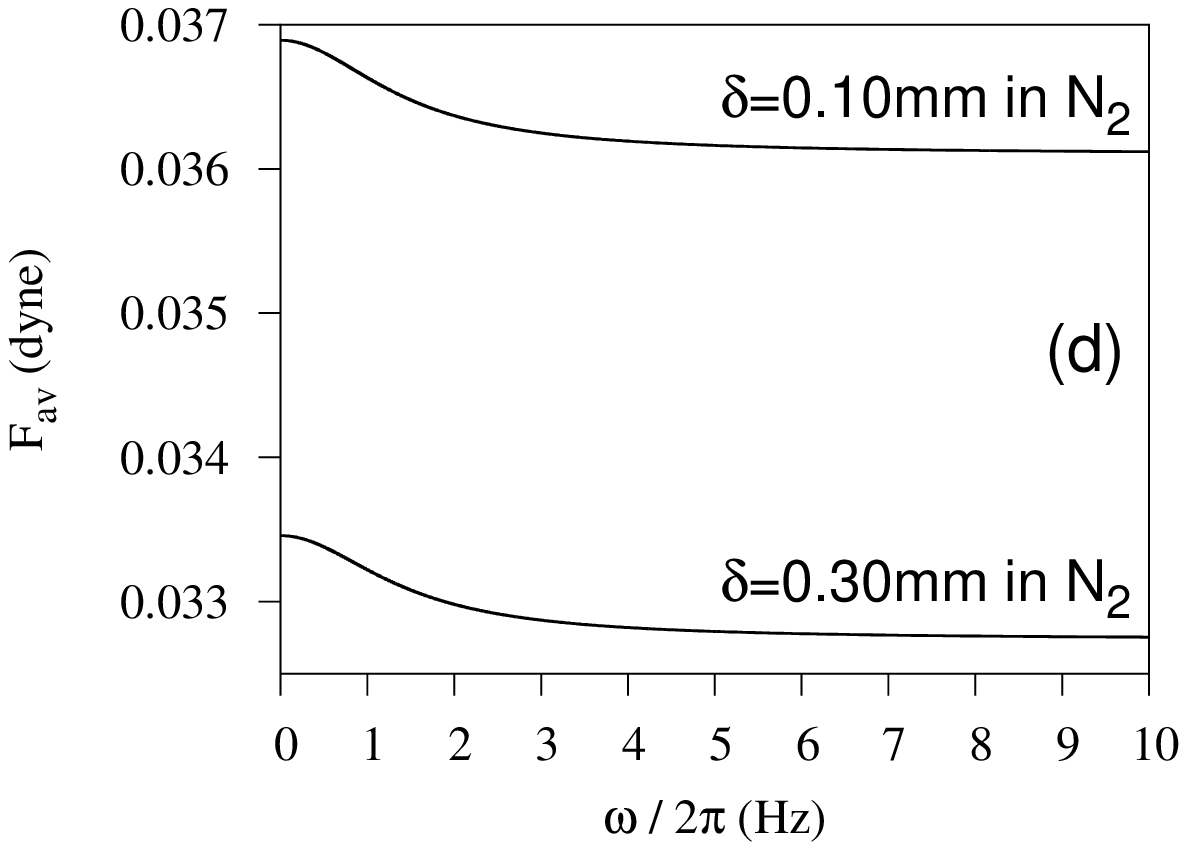, scale=0.6}
    \end{center}
  \end{minipage}
\end{tabular}
\caption{\label{fig:sn2} \footnotesize{The radial component of the
time-averaged force between two identical spheres of
$\mathrm{SrTiO_{3}}$ separated by $R$, plotted as a function of
frequency for host materials consisting of silicone oil [(a) and
(b)] and $N_{2}$ [(c) and (d)], with gap spacings
$\delta=0.10\,\mathrm{mm}$ and $\delta=0.30\,\mathrm{mm}$. The
applied electric field is $E_{0}=71.3\,\mathrm{V\!/mm}$ and
$a=3.15\,\mathrm{mm}$ for all the cases. The electric field is
parallel to the line between two spheres in (a) and (c) and
perpendicular to that line in (b) and (d).}}
\end{figure}

\noindent {\footnotesize TABLE II.  Parameters for the
calculations shown in Figs.\ \ref{fig:sc}--\ref{fig:sn2}.  The
columns denote the material, the real part of its dielectric
constant, and its conductivity (in S/m).  All except for SrTiO$_3$
are used as host materials in the suspensions.}
\newpage

\begin{center}
FIGURES
\end{center}

\vspace{0.1in}

%


\end{document}